\documentclass[pra,twocolumn,superscriptaddress,aps,longbibliography]{revtex4-1}%
\usepackage{amsfonts}
\usepackage{mathtools}
\usepackage{amsmath}
\usepackage{amssymb}
\usepackage{dsfont}
\usepackage{graphicx}
\usepackage{xcolor}
\usepackage{enumitem}
\usepackage{physics}

\usepackage[colorlinks=true,linkcolor=blue,citecolor=red,plainpages=false,pdfpagelabels]
 {hyperref}
\usepackage{cleveref}
\usepackage{color}
\definecolor{dgreen}{HTML}{006600}
\usepackage{times}

\providecommand{\U}[1]{\protect\rule{.1in}{.1in}}
\newtheorem{theorem}{Theorem}

\newtheorem{corollary}{Corollary}

\def\Tr{\operatorname{Tr}}

\def\kex{\operatorname{EXT}_{k}}

\def\supp{\operatorname{supp}}

\def\>{\rangle}
\def\<{\langle}
\def\V{\Vert}

\newcommand{\mc}[1]{\mathcal{#1}}

\newcommand{\tf}[1]{\mathbf{#1}}

\begin{document}
\preprint{ }
\title[]{Extendibility limits the performance of quantum processors}

\author{Eneet Kaur}
\affiliation{Hearne Institute for Theoretical Physics, Department of Physics and Astronomy,
Louisiana State University, Baton Rouge, Louisiana 70803, USA}
\author{Siddhartha Das}
\affiliation{Hearne Institute for Theoretical Physics, Department of Physics and Astronomy,
Louisiana State University, Baton Rouge, Louisiana 70803, USA}
\author{Mark M. Wilde}
\affiliation{Hearne Institute for Theoretical Physics, Department of Physics and Astronomy,
Louisiana State University, Baton Rouge, Louisiana 70803, USA}
\affiliation{Center for Computation and Technology, Louisiana State University, Baton
Rouge, Louisiana 70803, USA}
\author{Andreas Winter}
\affiliation{ICREA \&{} F\'{\i}sica 
 Te\`{o}rica: Informaci\'{o} i Fen\`{o}mens Qu\`{a}ntics, 
 Departament de F\'{\i}sica, Universitat Aut\`{o}noma de Barcelona, 
 ES-08193 Bellaterra (Barcelona), Spain}
\pacs{PACS number}

\begin{abstract}
Resource theories in quantum information science are helpful for the study and quantification of the performance of information-processing tasks that involve quantum systems. These resource theories also find applications in other areas of study; e.g., the resource theories of entanglement and coherence have found use and implications in the study of quantum thermodynamics and memory effects in quantum dynamics. In this paper, we introduce the resource theory of unextendibility, which is associated to the inability of extending quantum entanglement in a given quantum state to multiple parties. The free states in this resource theory are the $k$-extendible states, and the free channels are $k$-extendible channels, which preserve the class of $k$-extendible states. We make use of this resource theory to derive non-asymptotic, upper bounds on the rate at which quantum communication or entanglement preservation is possible by utilizing an arbitrary quantum channel a  finite number of times, along with the assistance of $k$-extendible channels at no cost. We then show that the bounds obtained are significantly tighter than previously known bounds for quantum communication over both the depolarizing and erasure channels. 

\end{abstract}
\volumeyear{year}
\volumenumber{number}
\issuenumber{number}
\eid{identifier}
\date{\today}
\startpage{1}
\endpage{10}
\maketitle

\textit{Introduction}---Recent years have seen progress in the development of programmable quantum computers and information processing devices; several  groups are actively developing superconducting quantum processors~\cite{GCS17} and satellite-to-ground quantum key distribution \cite{LCL+17}. It is thus pertinent to establish benchmarks on the information-processing capabilities of quantum devices that are able to process a finite number of qubits reliably. Experimentalists can then employ these benchmarks to evaluate how far they are from achieving the fundamental limitations on performance. 

In this paper, we first develop a resource theory of unextendibility and then apply it to bound the performance of
quantum processors. In particular, the resource theory of unextendibility leads to non-asymptotic upper bounds on the rate at which entanglement can be preserved when using a given quantum channel a  finite number of times. We then apply this general bound to  depolarizing and erasure channels, which are common models of noise in quantum processors. For these channels, we find that our bounds are significantly tighter than previously known non-asymptotic bounds from \cite{TBR15,WFD17}. 

The resource theory of unextendibility can be understood as a relaxation of the well known resource
theory of entanglement \cite{BDSW96,HHHH09}, and it is a relaxation alternative to the resource theory of negative partial transpose\ states from
\cite{Rai99,Rai01}, in which the free states are the positive partial
transpose (PPT) states and the free channels are completely PPT-preserving channels. In the resource theory of entanglement, the free states are the separable
states, those not having any entanglement at all. 
Any separable state $\sigma_{AB}$ can be written as
$
\sigma_{AB}%
=\sum_{x}p(x)\tau_{A}^{x}\otimes\omega_{B}^{x}$,
 where $p(x)$ is a probability
distribution and $\{\tau_{A}^{x}\}_{x}$ and $\{\omega_{B}^{x}\}_{x}$ are sets
of states; the free channels are those that can be performed by local
operations and classical communication (LOCC)
\cite{BDSW96,CLM+14}. An LOCC channel $\mathcal{L}_{AB\rightarrow A^{\prime
}B^{\prime}}$ is a separable super-operator
 (although the converse is not true), and can hence  be written as
$
\mathcal{L}_{AB\rightarrow A^{\prime}B^{\prime}}=\sum_{y}\mathcal{E}%
_{A\rightarrow A^{\prime}}^{y}\otimes\mathcal{F}_{B\rightarrow B^{\prime}}%
^{y},
$
where $\{\mathcal{E}_{A\rightarrow A^{\prime}}^{y}\}_{y}$ and $\{\mathcal{F}%
_{B\rightarrow B^{\prime}}^{y}\}_{y}$ are sets of completely positive (CP)
maps such that $\mathcal{L}_{AB\rightarrow A^{\prime}B^{\prime}}$ is trace
preserving. A special kind of LOCC channel is a one-way (1W-) LOCC channel from $A$
to $B$, in which Alice performs a quantum instrument, sends the classical
outcome to Bob, who then performs a quantum channel conditioned on the
classical outcome received from Alice. As such, any 1W-LOCC channel takes the
form stated above, except that $\{\mathcal{E}_{A\rightarrow
A^{\prime}}^{y}\}_{y}$ is a set of CP\ maps such that the sum map $\sum
_{y}\mathcal{E}_{A\rightarrow A^{\prime}}^{y}$ is trace preserving, while
$\{\mathcal{F}_{B\rightarrow B^{\prime}}^{y}\}_{y}$ is a set of quantum channels.

The set of free states in the resource theory of unextendibility is larger than the set of free states in the resource theory of entanglement. By relaxing the resource theory of entanglement in this way, we obtain tighter, non-asymptotic bounds on the entanglement transmission rates of a quantum channel.   

Before we begin with our development, we note here that detailed proofs of all statements that follow are given in our companion paper \cite{KDWW18}. 

\textit{Resource theory of unextendibility}---In  the resource theory of unextendibility, there is
implicitly a positive integer $k\geq2$, with respect to which the theory is
defined. The free states in this resource theory are the $k$-extendible states
\cite{W89a,DPS02,DPS04}, a prominent notion in quantum information and
entanglement theory that we recall now. For a positive integer $k\geq2$, a
bipartite state $\rho_{AB}$ is $k$-extendible with respect to system $B$  if

\begin{enumerate}[wide, labelwidth=!, labelindent=0pt]
\item \textbf{(State Extension)} There exists a state $\omega
_{AB_{1}\cdots B_{k}}$ that extends $\rho_{AB}$, so that
$
\operatorname{Tr}_{B_{2}\cdots B_{k}}\{\omega_{AB_{1}\cdots B_{k}}\}=\rho_{AB},%
$
with systems $B_{1}$ through $B_{k}$ each isomorphic to system $B$ of
$\rho_{AB}$.

\item \textbf{(Permutation Invariance)} The extension state $\omega
_{AB_{1}\cdots B_{k}}$ is invariant with respect to permutations of the $B$
systems, in the sense that
$
\omega_{AB_{1}\cdots B_{k}}=W_{B_{1}\cdots B_{k}}^{\pi}\omega_{AB_{1}\cdots
B_{k}}W_{B_{1}\cdots B_{k}}^{\pi\dag},
$
where $W_{B_{1}\cdots B_{k}}^{\pi}$ is a unitary representation of the
permutation $\pi\in S_{k}$, with $S_{k}$ denoting the symmetric group.
\end{enumerate}


To give some physical context to the definition of a $k$-extendible state, suppose that Alice and Bob share a bipartite state and that Bob subsequently mixes his system and the vacuum state at a 50:50 beamsplitter. Then the resulting state of Alice's system and one of the outputs of the beamsplitter is a two-extendible state by construction. As a generalization of this, suppose that Bob sends his system through the $N$-splitter of \cite[Eq.~(10)]{vLB00}, with the other input ports set to the vacuum state. Then the state of Alice's system and one of the outputs of the $N$-splitter is $N$-extendible by construction. One could also physically realize $k$-extendible states in a similar way by means of approximate quantum cloning machines \cite{RevModPhys.77.1225}.

It is worthwhile to mention that there are free states in the resource theory of unextendibility that are not free in the resource theory of entanglement. For example, if we send one share of the maximally entangled state $\Phi_{AB}$ through a 50\% erasure channel \cite{GBP97}, then the resulting state $\frac{1}{2}(\Phi_{AB} + I_A/2 \otimes \vert e\rangle \langle e \vert_B)$ is a two-extendible state, and is thus free in the resource theory of unextendibility for $k=2$. However, this state has distillable entanglement via LOCC \cite{PhysRevLett.78.3217}, and so it is not free in the resource theory of entanglement.

Let $\operatorname{EXT}_{k}(A;B)$ denote the set of $k$-extendible
states, where with this notation and as above, we take it as implicit that the system $B$ is being extended. The $k$-extendible states are a
relaxation of the set of separable (unentangled) states, in the sense that a
separable state is $k$-extendible for any positive integer $k\geq2$.
Furthermore, if a state $\rho_{AB}$\ is entangled, then there exists some~$k$
for which $\rho_{AB}$ is not $k$-extendible, and $\rho_{AB}$ is not
$\ell$-extendible for all $\ell\geq k$ \cite{DPS02,DPS04}. 

We define the free channels in the resource theory of unextendibility to be bipartite channels that satisfy two constraints
 generalizing those given above for the free states. Recall that a bipartite channel $\mathcal{N}%
_{AB\rightarrow A^{\prime}B^{\prime}}$ has two input systems $A$ and $B$ and two output systems $A'$ and $B'$. The systems $A$ and $A'$ are held by a single party Alice, and the systems $B$ and $B'$ are held by another party Bob. It could be the case that any of these systems encompass a number of smaller subsystems, and we make use of this in what follows. We define a bipartite channel $\mathcal{N}%
_{AB\rightarrow A^{\prime}B^{\prime}}$ to be $k$-extendible if

\begin{enumerate}[wide, labelwidth=!, labelindent=0pt]
\item \textbf{(Channel Extension)}\ There exists a quantum channel
$\mathcal{M}_{AB_{1}\cdots B_{k}\rightarrow A^{\prime}B_{1}^{\prime}\cdots
B_{k}^{\prime}}$ that extends $\mathcal{N}_{AB\rightarrow A^{\prime}B^{\prime
}}$, in the sense that the following equality holds for all quantum states
$\theta_{AB_{1}\cdots B_{k}}$:
$
\operatorname{Tr}_{B_{2}^{\prime}\cdots B_{k}^{\prime}}\{\mathcal{M}%
_{AB_{1}\cdots B_{k}\rightarrow A^{\prime}B_{1}^{\prime}\cdots B_{k}^{\prime}%
}(\theta_{AB_{1}\cdots B_{k}})\}
=\mathcal{N}_{AB\rightarrow A^{\prime}B^{\prime}}(\theta_{AB_{1}})$,
with $B_{1}\cdots B_{k}$ each isomorphic to $B$,
and $B_{1}^{\prime}\cdots B_{k}^{\prime}$ each isomorphic to $B^{\prime}$.

\item \textbf{(Permutation Covariance)} The extension channel $\mathcal{M}%
_{AB_{1}\cdots B_{k}\rightarrow A^{\prime}B_{1}^{\prime}\cdots B_{k}^{\prime}%
}$ is covariant with respect to permutations of the input $B$ and output
$B^{\prime}$ systems, in the sense that the following equality holds for all
quantum states $\theta_{AB_{1}\cdots B_{k}}$:
$
\mathcal{M}_{AB_{1}\cdots B_{k}\rightarrow A^{\prime}B_{1}^{\prime}\cdots
B_{k}^{\prime}}(W_{B_{1}\cdots B_{k}}^{\pi}\theta_{AB_{1}\cdots B_{k}}%
W_{B_{1}\cdots B_{k}}^{\pi\dag})=
W_{B_{1}^{\prime}\cdots B_{k}^{\prime}}^{\pi}\mathcal{M}_{AB_{1}\cdots
B_{k}\rightarrow A^{\prime}B_{1}^{\prime}\cdots B_{k}^{\prime}}(\theta
_{AB_{1}\cdots B_{k}})W_{B_{1}^{\prime}\cdots B_{k}^{\prime}}^{\pi\dag}$,
where $W_{B_{1}\cdots B_{k}}^{\pi}$ and $W_{B_{1}^{\prime}\cdots B_{k}%
^{\prime}}^{\pi}$ are unitary representations of the permutation $\pi\in
S_{k}$.
\end{enumerate}

\noindent The first condition above can be understood as a
no-signaling condition. That is, it implies that it is impossible for the parties
controlling the $B_{2}\cdots B_{k}$ systems to communicate to the parties
holding systems $A^{\prime}B_{1}^{\prime}$. 

 We advocate that our definition above is a natural channel generalization of
state extendibility, since the reduced channel $\mathcal{N}_{AB\rightarrow
A^{\prime}B^{\prime}}$ of the channel extension $\mathcal{M}_{AB_{1}\cdots
B_{k}\rightarrow A^{\prime}B_{1}^{\prime}\cdots B_{k}^{\prime}}$ is defined in
an unambiguous way only when we impose a no-signaling constraint. Furthermore,
the above definition is quite natural in the resource theory of
unextendibility developed here, as evidenced by the following theorem:

\begin{theorem}\label{Thm-1}
Let $\rho_{AB}\in\kex(A;B)$, and let $\mathcal{N}_{AB\rightarrow
A^{\prime}B^{\prime}}$ be a $k$-extendible channel. Then the output state
$\mathcal{N}_{AB\rightarrow A^{\prime}B^{\prime}}(\rho_{AB})$ is $k$-extendible.
\end{theorem}

The above theorem is fundamental for the resource theory of unextendibility, indicating that the $k$-extendible channels are free, as they preserve the free states.  

There are several interesting classes of $k$-extendible channels that we can consider. Even if it might seem trivial, we should mention that a particular
kind of $k$-extendible channel is in fact a $k$-extendible state, in which the
input systems $A$ and $B$ are trivial. Thus, $k$-extendible channels can generate $k$-extendible states.

Any 1W-LOCC channel is $k$-extendible for all $k\geq2$, similar to
the way in which any separable state is $k$-extendible for all $k\geq2$. Thus,
a 1W-LOCC channel is free in the resource theory of unextendibility. The fact that a 1W-LOCC channel takes a $k$-extendible input state
to a $k$-extendible output state had already been observed for the
special case $k=2$ in \cite{NH09}.

\textit{Quantifying unextendibility}---In any resource theory, it is pertinent to quantify the resourcefulness of the resource states and channels. It is desirable for any quantifier to be non-negative, attain its minimum for the free states and channels, and be monotone under the action of a free channel \cite{BG15}. With this in mind, we define the $k$-unextendible generalized divergence of an arbitrary density operator $\rho_{AB}$ as follows:
\begin{equation}
\tf{E}_{k}(A;B)_{\rho}=\inf_{\sigma_{AB}\in\kex(A;B)}\tf{D}(\rho_{AB}\Vert\sigma_{AB}), \label{eq:gen-div-unext}
\end{equation}
where $\tf{D}(\rho\Vert \sigma)$ denotes a generalized divergence  \cite{PV10,SW12}, which is any  quantifier of the distinguishability of states $\rho$ and $\sigma$ that is monotone under the action of a quantum channel. Special cases of the quantifier in \eqref{eq:gen-div-unext} were previously defined in \cite{NH09,B08} (relative entropy to two-extendible states and to $k$-extendible states, respectively), \cite{PhysRevA.74.052301} (best two-extendible approximation, related to max-relative entropy of unextendibility defined here), and \cite{HMW13} (maximum $k$-extendible fidelity). 

Particular examples of generalized divergences between states $\rho$ and $\sigma$ are the $\varepsilon$-hypothesis-testing divergence $D^\varepsilon_h(\rho\Vert\sigma)$ \cite{BD10,WR12}, and the max-relative entropy $D_{\max}(\rho\Vert\sigma)$ \cite{D09,Dat09}, where for $\varepsilon\in[0,1]$,
\begin{equation*}
 D^\varepsilon_h\!\left(\rho\Vert\sigma\right) \coloneqq  -\log_2\inf_{\Lambda\in[0,I] }\{\Tr\{\Lambda\sigma\}:
 \Tr\{\Lambda\rho\}\geq 1-\varepsilon\},
\end{equation*}
 and 
$
D_{\max}(\rho\V\sigma)\coloneqq \inf\{\lambda:\ \rho \leq 2^\lambda\sigma\}
$
in the case that $\supp(\rho)\subseteq\supp(\sigma)$, and otherwise $D_{\max}(\rho\V\sigma)=+\infty$.

\textit{Information-processing tasks}---Now that we have established the free
states and channels in the resource theory of unextendibility, we are ready to
discuss tasks that can be performed in it. We consider two
main tasks here:\ entanglement distillation and quantum communication
with the assistance of $k$-extendible channels. The goal of these protocols is
to use many copies of a bipartite state or many invocations of a quantum
channel, along with the free assistance of $k$-extendible channels, in order
to generate a high-fidelity maximally entangled state with as much
entanglement as possible. This kind of task was defined and developed in
\cite{SSW08}, albeit with the assistance of a particular kind of
$k$-extendible channel and only the case $k=2$ was considered there, generalizing the usual notion of entanglement distillation and quantum communication protocols from \cite{BDSW96,PhysRevA.54.2614,PhysRevA.54.2629,L97,BNS98,BKN98,Sho02,D05}.

Let $n,M\in\mathbb{Z}^{+}$ and $\varepsilon\in\left[  0,1\right]  $. Let
$\rho_{AB}$ be a bipartite state. An $\left(  n,M,\varepsilon\right)  $
entanglement distillation protocol assisted by $k$-extendible channels begins
with Alice and Bob sharing $n$ copies of $\rho_{AB}$, to which they apply a
$k$-extendible channel $\mathcal{K}_{A^{n}B^{n}\rightarrow M_{A}M_{B}}$, where it is understood that this is a bipartite channel with Alice possessing systems $A^n$ and $M_A$ and Bob possessing systems $B^n$ and $M_B$. The
resulting state satisfies the following performance condition:%
\begin{equation}
F(\mathcal{K}_{A^{n}B^{n}\rightarrow M_{A}M_{B}}(\rho_{AB}^{\otimes n}%
),\Phi_{M_{A}M_{B}})\geq1-\varepsilon, \label{eq:ent-dist-criterion}
\end{equation}
where $\Phi_{M_{A}M_{B}}\coloneqq \frac{1}{M}\sum_{m,m^{\prime}}|m\rangle\langle
m^{\prime}|_{M_{A}}\otimes|m\rangle\langle m^{\prime}|_{M_{B}}$ is a maximally
entangled state of Schmidt rank $M$ and $F(\omega,\tau)\coloneqq\left\Vert
\sqrt{\omega}\sqrt{\tau}\right\Vert _{1}^{2}$ is the quantum fidelity
\cite{U76}. Let $D^{(k)}(\rho_{AB},n,\varepsilon)$ denote the non-asymptotic
distillable entanglement with the assistance of $k$-extendible channels; i.e.,
$D^{(k)}(\rho_{AB},n,\varepsilon)$ is equal to the maximum value of $\frac
{1}{n}\log_{2}M$ such that there exists an $(n,M,\varepsilon)$ protocol for
$\rho_{AB}$ satisfying \eqref{eq:ent-dist-criterion}.

We define two different variations of quantum communication, with one simpler
and one more involved. Let $\mathcal{N}_{A\rightarrow B}$ denote a quantum
channel. In the simpler version, an $\left(  n,M,\varepsilon\right)  $ entanglement transmission protocol assisted by a $k$-extendible post-processing begins with
Alice preparing a maximally entangled state $\Phi_{RA'}$ of Schmidt rank $M$. She applies a quantum channel $\mathcal{E}_{A' \to A^n}$, which serves as an encoding and leads to a state $\rho_{RA^{n}}
\coloneqq \mathcal{E}_{A' \to A^n}(\Phi_{RA'})$. She transmits the systems
$A^{n}\coloneqq A_{1}\cdots A_{n}$ using the channel $\mathcal{N}_{A\rightarrow
B}^{\otimes n}$. Alice and Bob then perform a $k$-extendible channel
$\mathcal{K}_{RB^{n}\rightarrow M_{A}M_{B}}$, such that
\begin{equation}
F(\mathcal{K}_{RB^{n}\rightarrow M_{A}M_{B}}(\mathcal{N}_{A\rightarrow
B}^{\otimes n}(\rho_{RA^{n}})),\Phi_{M_{A}M_{B}})\geq1-\varepsilon.
\label{eq:ent-trans-criterion}
\end{equation}
Let $Q_{\text{I}}^{(k)}(\mathcal{N}_{A\rightarrow B},n,\varepsilon)$ denote
the non-asymptotic quantum capacity assisted by a
$k$-extendible post-processing; i.e., $Q_{\text{I}}^{(k)}(\mathcal{N}_{A\rightarrow
B},n,\varepsilon)$ is the maximum value of $\frac{1}{n}\log_{2}M$ such that
there exists an $(n,M,\varepsilon)$ protocol for $\mathcal{N}_{A\rightarrow
B}$ satisfying~\eqref{eq:ent-trans-criterion}.

For the cases of entanglement distillation and the simpler version of entanglement transmission, note that an $(n,M,\varepsilon)$ entanglement distillation protocol for the state $\rho_{AB}$ is a $(1,M,\varepsilon)$ protocol for the state $\rho_{AB}^{\otimes n}$ and vice versa. Similarly, an $(n,M,\varepsilon)$ entanglement transmission protocol for the channel $\mathcal{N}_{A \to B}$ is a $(1,M,\varepsilon)$ protocol for the channel $\mathcal{N}_{A \to B}^{\otimes n}$ and vice versa.

In the more involved version of entanglement transmission, every channel use is interleaved with a
$k$-extendible channel, similar to the protocols considered in
\cite{TGW14,TGW14nat,KW17a}. Specifically, the protocol is a special case of
one discussed in \cite{KW17a} for general resource theories.
We do not discuss
these protocols in detail here, but we simply note that, for an
$(n,M,\varepsilon)$ quantum communication protocol assisted by $k$-extendible
channels, the performance criterion is that the final state of the protocol
should have fidelity $\geq1-\varepsilon$ to a maximally entangled state
$\Phi_{M_{A}M_{B}}$ of Schmidt rank~$M$. Let $Q_{\text{II}}^{(k)}%
(\mathcal{N}_{A\rightarrow B},n,\varepsilon)$ denote the non-asymptotic
quantum capacity assisted by $k$-extendible channels; i.e.,
$Q_{\text{II}}^{(k)}(\mathcal{N}_{A\rightarrow B},n,\varepsilon)$ is the
maximum value of $\frac{1}{n}\log_{2}M$ such that there exists an
$(n,M,\varepsilon)$ protocol for $\mathcal{N}_{A\rightarrow B}$ as described for the more involved case above.

\begin{theorem}\label{thm:q-com-converse}
The following bound holds for all  $ k\geq 2$ and for  any $(1,M,\varepsilon)$ entanglement transmission protocol that uses a  channel $\mathcal{N}$ assisted by a $k$-extendible post-processing:
\begin{equation}
-\log_2\!\left[\frac{1}{M}+\frac{1}{k}-\frac{1}{M k}\right]\leq \sup_{\psi_{RA}} E^\varepsilon_{k}(R;B)_{\tau}, 
\end{equation}
where 
$
E^{\varepsilon}_{k}(R;B)_{\rho}\coloneqq \inf_{\sigma_{RB}\in\kex(R;B)}D^\varepsilon_h(\rho_{RB}\Vert\sigma_{RB})$,
 $\tau_{RB}\coloneqq \mc{N}_{A\to B}(\psi_{RA})$, and the optimization is with respect to pure states $\psi_{RA}$ such that $R\simeq A$.
 The following bound holds for all  $ k\geq 2$ and for  any $(1,M,\varepsilon)$ entanglement distillation protocol that uses a quantum state $\rho_{AB}$ assisted by a $k$-extendible post-processing:
\begin{equation}
-\log_2\!\left[\frac{1}{M}+\frac{1}{k}-\frac{1}{M k}\right]\leq E^\varepsilon_{k}(A;B)_{\rho}. 
\end{equation}
\end{theorem}

The proof of the above theorem follows by employing the fact that $E^{\varepsilon}_{k}$ does not increase under the action of a $k$-extendible channel, because the extendibility of a $k$-extendible state does not change under the action of $U\otimes U^\ast$ for a unitary $U$, and by employing
\cite[Theorem III.8]{JV13}.
\begin{theorem} \label{thm:q-cap-bnd-interleaved}
The following bound holds for all  $ k\geq 2$ and for  any $(n,M,\varepsilon)$  quantum communication protocol employing $n$ uses of a channel $\mathcal{N}$ interleaved by $k$-extendible channels:
\begin{equation*}
-\log_2\!\left[\frac{1}{M}+\frac{1}{k}-\frac{1}{M k}\right]  
 \leq n E^{\max}_{k}(\mathcal{N})+ \log_2\!\left(\frac{1}{1-\varepsilon}\right), 
\end{equation*}
where $$E^{\max}_{k}(\mathcal{N}) \coloneqq \sup_{\psi_{RA}} \inf_{\sigma_{RB}\in\kex(R;B)}D_{\max}(\tau_{RB}\Vert\sigma_{RB}),$$ 
 $\tau_{RB}\coloneqq \mc{N}_{A\to B}(\psi_{RA})$, and the optimization is with respect to pure states $\psi_{RA}$ with $|R|=|A|$.
\end{theorem}

We note here that special cases of the entanglement distillation and quantum
communication protocols described above occur when the $k$-extendible
assisting channels are taken to be 1W-LOCC\ channels. As such,
$D^{(k)}(\rho_{AB},n,\varepsilon)$, $Q_{\text{I}}^{(k)}(\mathcal{N}%
_{A\rightarrow B},n,\varepsilon)$, and $Q_{\text{II}}^{(k)}(\mathcal{N}%
_{A\rightarrow B},n,\varepsilon)$ are upper bounds on the non-asymptotic
distillable entanglement and capacities when 1W-LOCC\ channels are
available for assistance.

\textit{Pretty strong converse for antidegradable channels}---As a direct application of Theorem~\ref{thm:q-cap-bnd-interleaved}, we revisit the ``pretty strong converse'' of \cite{MW13} for antidegradable channels. Recall that a channel $\mathcal{N}_{A \to B}$ is antidegradable \cite{CG06,M10} if the output state $\mathcal{N}_{A \to B}(\rho_{RA})$ is two-extendible for any input state $\rho_{RA}$. Due to this property, antidegradable channels have zero asymptotic quantum capacity \cite{PhysRevLett.78.3217,Holevo2008}. Theorem~\ref{thm:q-cap-bnd-interleaved} implies the following bound for the non-asymptotic case:
\begin{corollary}
Fix $\varepsilon \in [0,1/2)$. The following bound holds for  any $(n,M,\varepsilon)$  quantum communication protocol employing $n$ uses of an antidegradable channel $\mathcal{N}$ interleaved by two-extendible channels:
$
\frac{1}{n}\log_2 M
 \leq \frac{1}{n}  \log_2\!\left(\frac{1}{1-2\varepsilon}\right). 
$
\end{corollary}

We conclude from the above inequality  that, for an antidegradable channel, there is a strong limitation on its ability to generate entanglement whenever the error parameter $\varepsilon < \tfrac12$, as is usually desired for applications in quantum computation. We also remark that the bound above is tighter than related bounds given in \cite{MW13}, and furthermore, the bound applies to quantum communication protocols assisted by interleaved two-extendible channels, which were not considered in~\cite{MW13}.

\textit{Limitations on
quantum devices}---In practice, the evolution effected by quantum processors is never a perfect unitary process. There is always some undesirable interaction with the environment, the latter of which is inaccessible to the processor. Furthermore, there are practical limitations on the ability to construct perfect unitary gates \cite{CCG+11}. The depolarizing  and erasure channels are two classes of noisy models for qubit quantum processors that are widely considered  (see \cite{LMR+17,BS10,DKP15}). 

Both families of channels mentioned above are covariant channels \cite{H02}; i.e., these channels are covariant with respect to a group $G$ with representations given by a unitary one-design. Thus, these channels can be simulated using 1W-LOCC with the Choi states as the resource states \cite[Section~VII]{CDP09}. Using this symmetry and the monotonicity of the unextendible generalized divergence under 1W-LOCC, we conclude that the optimal input state to a covariant channel $\mc{N}$, with respect to the upper bound in Theorem~\ref{thm:q-com-converse}, is a maximally entangled state~$\Phi_{RA}$. Also, for any $(n,M,\varepsilon)$ quantum communication protocol conducted over a covariant channel and assisted by any $k$-extendible channel, the optimal input state  is~$\Phi_{RA}^{\otimes n}$ and $Q_{\text{II}}^{(k)}(\mathcal{N}%
_{A\rightarrow B},n,\varepsilon) = Q_{\text{I}}^{(k)}(\mathcal{N}
_{A\rightarrow B},n,\varepsilon)$; i.e., an upper bound on non-asymptotic quantum capacity $Q^{(k)}_{\text{II}}$ is given by Theorem~\ref{thm:q-com-converse}. 

A qubit depolarizing channel acts on an input state $\rho$ as 
$
\mathcal{D}^p_{A\to B}(\rho) = (1-p)\rho + \frac{p}{3}(X\rho X+Y\rho Y + Z \rho Z),
$
where $p\in [0,1]$ is the depolarizing parameter, and $X$, $Y$, and $Z$ are Pauli operators. The best known upper bound on the aysmptotic quantum capacity of this channel for values of $p\in[0,\tfrac14)$ was recently derived in \cite{LDS17,LLS17}, and this channel has zero asymptotic quantum capacity for $p\in[\tfrac14,1]$ \cite{BDEFMS98,C00}. 

\begin{figure}
\centering
\includegraphics[width=0.7\linewidth]{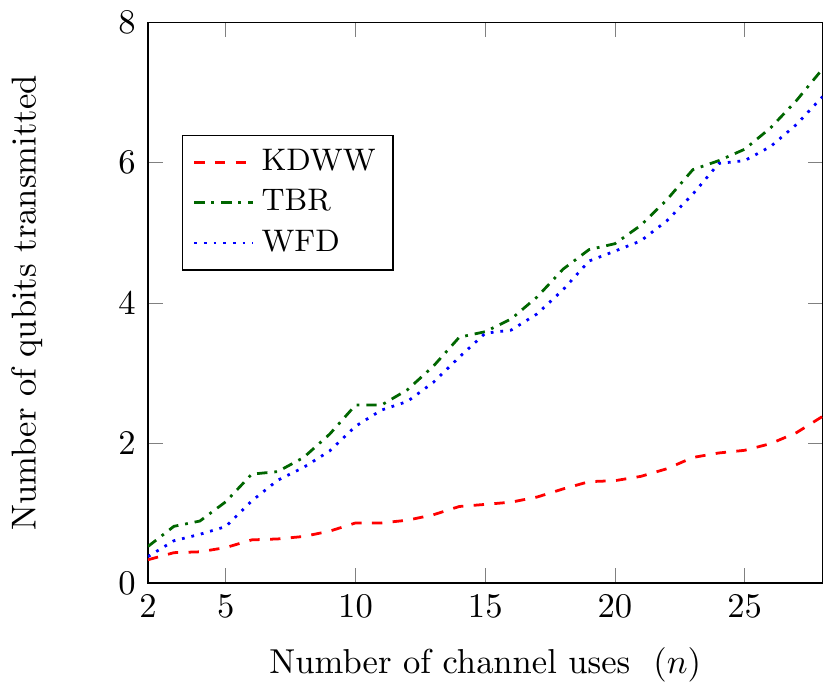}
\caption{Upper bounds on the number of qubits that can be reliably transmitted over a depolarizing channel with $p=0.15$, and $\varepsilon= 0.05$. The red dashed line is  from Theorem \ref{thm:q-com-converse}. The green dash-dotted and blue dotted lines are upper bounds from \cite{TBR15} and \cite{WFD17}, respectively.}
\label{fig:dep2}
\end{figure}

With the goal of bounding the non-asymptotic quantum capacity of $\mathcal{D}^p$,  we make a particular choice of the $k$-extendible state for $E_k^\varepsilon$ (which need not be optimal) to be a tensor power of the isotropic states $\sigma_{AB}^{(t,2)}$, which is similar to what was done in \cite{TBR15}. The inequality in Theorem~\ref{thm:q-com-converse} then reduces to
\begin{equation}\label{eq:bound-prl}
\frac{1}{n}\log_2 M \leq \frac{1}{n}\log_2 \!\left(1-\frac{1}{k}\right)
 - \frac{1}{n}\log_2\! \left(f(\varepsilon,p,t)-\frac{1}{k}\right),
\end{equation}
where $f(\varepsilon,p,t)=2^{-D^\varepsilon_h\left(\{1-p,p\}^{\otimes n}\left\Vert \{t,1-t\}^{\otimes n}\right)\right.}$ and $\{1-p,p\}$ denotes a Bernoulli distribution. 
The optimal measurement (Neyman-Pearson test) for the resulting hypothesis testing relative entropy  between Bernoulli distributions is then well known \cite{PPV10} (see also  \cite{MW14}), giving an explicit upper bound on the rate $\frac{1}{n}\log_2 M$. Figure~\ref{fig:dep2} compares various upper bounds on the number of qubits that can be reliably transmitted over $n$ uses of the depolarizing channel. The bounds plotted are the ones derived from Theorem~\ref{thm:q-com-converse} (labeled ``KDWW''), as well as two other known  upper bounds on non-asymptotic quantum capacities  \cite{TBR15,WFD17}. The figure demonstrates that the bounds coming from the resource theory of unextendibility are significantly tighter than those from \cite{TBR15,WFD17}. Note that \eqref{eq:bound-prl} converges to the upper bound from \cite{TBR15,WTB17} in the limit  $k \rightarrow \infty$.

\begin{figure}
\centering
\includegraphics[width=0.7\linewidth]{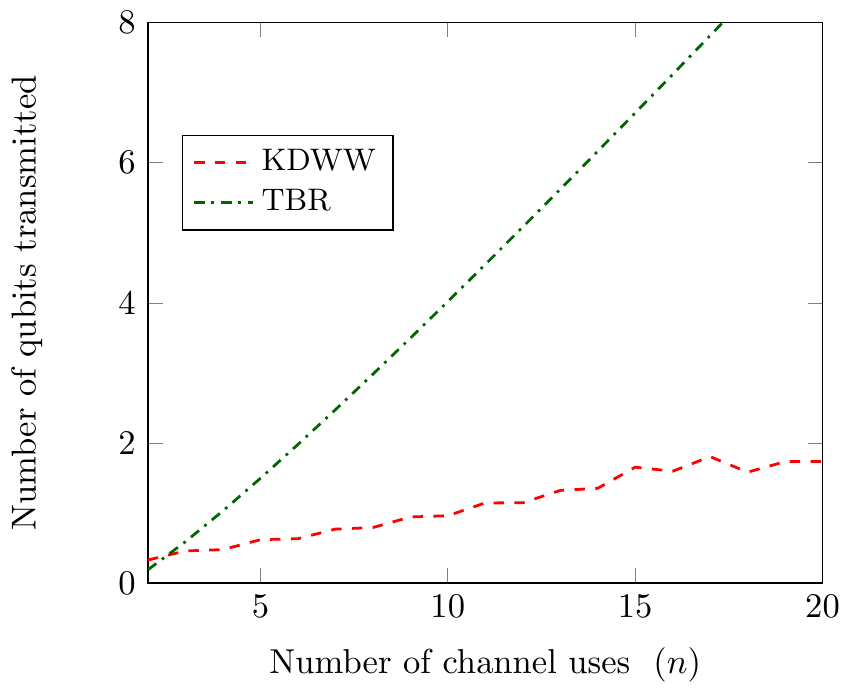}
\caption{Upper bounds on the number of qubits that can be reliably transmitted over an erasure channel with $p=0.35$, and $\varepsilon= 0.05$. The red dashed line is the bound from Theorem~\ref{thm:q-com-converse}. The green dash-dotted line is an upper bound from~\cite{TBR15}.}
\label{fig-erasure1}
\end{figure}

A qubit erasure channel acts on an input state $\rho$  as  
$
\mathcal{E}^p_{A\rightarrow B}(\rho_A) = (1-p)\rho_{B}+p \op{e}_B$ \cite{GBP97},
where $p\in[0,1]$ is the erasure probability, and the erasure state $\op{e}$ is orthogonal to the input Hilbert space. We employ the symmetries of the erasure channel to make a particular choice of the $k$-extendible state for $E_k^\varepsilon$. Theorem~\ref{thm:q-com-converse} gives upper bounds on the number of  qubits that can be reliably transmitted over $n$ uses of the erasure channel. The bounds that we obtain are not necessarily optimal, but they still are significantly tighter than those from \cite{TBR15}. See Figure~\ref{fig-erasure1}. 

\textit{Discussion}---In this paper, we  developed the resource theory of unextendibility and discussed limits that it places on the performance of finite-sized quantum processors. The free states in this resource theory are $k$-extendible states, and the free channels are the $k$-extendible channels. We determined non-asymptotic upper bounds on the rate at which qubits can be transmitted over a finite number of uses of a given quantum channel. The bounds coming from the resource theory of unextendibility are significantly tighter than those in \cite{TBR15,WFD17} for depolarizing and erasure channels.  

It would be interesting to explore the resource theory of unextendibility  further. One plausible direction would be to use this resource theory to obtain non-asymptotic converse bounds on the entanglement distillation rate of bipartite quantum interactions and compare with the bounds obtained in \cite{DBW17,BDW18}. Another direction is to analyze the bounds in Theorem~\ref{thm:q-com-converse} for other noise models that are practically relevant. Finally,  it remains open to link the bounds developed here with the open problem of finding a strong converse for the quantum capacity of degradable channels \cite{MW13}. To solve that problem, recall that one contribution of \cite{MW13} was to reduce the question of the strong converse of degradable channels to that of establishing the strong converse for symmetric channels.
 
\textit{Note}---We 
noticed the related work ``Optimising practical entanglement 
distillation'' by Rozpedek \textit{et al}.~recently posted as 
arXiv:1803.10111, which like us uses extendibility to address
entanglement distillation, and which presents results that are 
complementary to ours.

\begin{acknowledgments}
We thank Sumeet Khatri, Vishal Katariya, Felix Lediztky, and Stefano Mancini for insightful discussions. SD acknowledges support from the LSU Graduate School Economic Development Assistantship. EK and MMW acknowledge support from the US Office of Naval Research and the National Science Foundation under Grant No.~1350397. Andreas Winter acknowledges support from the ERC
Advanced Grant IRQUAT, the Spanish MINECO, projects FIS2013-40627-P and FIS2016-86681-P, with
the support of FEDER funds, and the Generalitat de Catalunya, CIRIT project 2014-SGR-966.
\end{acknowledgments}


\begin{thebibliography}{60}%
\makeatletter
\providecommand \@ifxundefined [1]{%
 \@ifx{#1\undefined}
}%
\providecommand \@ifnum [1]{%
 \ifnum #1\expandafter \@firstoftwo
 \else \expandafter \@secondoftwo
 \fi
}%
\providecommand \@ifx [1]{%
 \ifx #1\expandafter \@firstoftwo
 \else \expandafter \@secondoftwo
 \fi
}%
\providecommand \natexlab [1]{#1}%
\providecommand \enquote  [1]{``#1''}%
\providecommand \bibnamefont  [1]{#1}%
\providecommand \bibfnamefont [1]{#1}%
\providecommand \citenamefont [1]{#1}%
\providecommand \href@noop [0]{\@secondoftwo}%
\providecommand \href [0]{\begingroup \@sanitize@url \@href}%
\providecommand \@href[1]{\@@startlink{#1}\@@href}%
\providecommand \@@href[1]{\endgroup#1\@@endlink}%
\providecommand \@sanitize@url [0]{\catcode `\\12\catcode `\$12\catcode
  `\&12\catcode `\#12\catcode `\^12\catcode `\_12\catcode `\%12\relax}%
\providecommand \@@startlink[1]{}%
\providecommand \@@endlink[0]{}%
\providecommand \url  [0]{\begingroup\@sanitize@url \@url }%
\providecommand \@url [1]{\endgroup\@href {#1}{\urlprefix }}%
\providecommand \urlprefix  [0]{URL }%
\providecommand \Eprint [0]{\href }%
\providecommand \doibase [0]{http://dx.doi.org/}%
\providecommand \selectlanguage [0]{\@gobble}%
\providecommand \bibinfo  [0]{\@secondoftwo}%
\providecommand \bibfield  [0]{\@secondoftwo}%
\providecommand \translation [1]{[#1]}%
\providecommand \BibitemOpen [0]{}%
\providecommand \bibitemStop [0]{}%
\providecommand \bibitemNoStop [0]{.\EOS\space}%
\providecommand \EOS [0]{\spacefactor3000\relax}%
\providecommand \BibitemShut  [1]{\csname bibitem#1\endcsname}%
\let\auto@bib@innerbib\@empty
\bibitem [{\citenamefont {Gambetta}\ \emph {et~al.}(2017)\citenamefont
  {Gambetta}, \citenamefont {Chow},\ and\ \citenamefont {Steffen}}]{GCS17}%
  \BibitemOpen
  \bibfield  {author} {\bibinfo {author} {\bibfnamefont {Jay~M.}\ \bibnamefont
  {Gambetta}}, \bibinfo {author} {\bibfnamefont {Jerry~M.}\ \bibnamefont
  {Chow}}, \ and\ \bibinfo {author} {\bibfnamefont {Matthias}\ \bibnamefont
  {Steffen}},\ }\bibfield  {title} {\enquote {\bibinfo {title} {Building
  logical qubits in a superconducting quantum computing system},}\ }\href
  {\doibase 10.1038/s41534-016-0004-0} {\bibfield  {journal} {\bibinfo
  {journal} {npj Quantum Information}\ }\textbf {\bibinfo {volume} {3}},\
  \bibinfo {pages} {2} (\bibinfo {year} {2017})}\BibitemShut {NoStop}%
\bibitem [{\citenamefont {Liao}\ \emph {et~al.}(2017)\citenamefont {Liao},
  \citenamefont {Cai}, \citenamefont {Liu}, \citenamefont {Zhang},
  \citenamefont {Li}, \citenamefont {Ren}, \citenamefont {Yin}, \citenamefont
  {Shen}, \citenamefont {Cao}, \citenamefont {Li}, \citenamefont {Li},
  \citenamefont {Chen}, \citenamefont {Sun}, \citenamefont {Jia}, \citenamefont
  {Wu}, \citenamefont {Jiang}, \citenamefont {Wang}, \citenamefont {Huang},
  \citenamefont {Wang}, \citenamefont {Zhou}, \citenamefont {Deng},
  \citenamefont {Xi}, \citenamefont {Ma}, \citenamefont {Hu}, \citenamefont
  {Zhang}, \citenamefont {Chen}, \citenamefont {Liu}, \citenamefont {Wang},
  \citenamefont {Zhu}, \citenamefont {Lu}, \citenamefont {Shu}, \citenamefont
  {Peng}, \citenamefont {Wang},\ and\ \citenamefont {Pan}}]{LCL+17}%
  \BibitemOpen
  \bibfield  {author} {\bibinfo {author} {\bibfnamefont {Sheng-Kai}\
  \bibnamefont {Liao}}, \bibinfo {author} {\bibfnamefont {Wen-Qi}\ \bibnamefont
  {Cai}}, \bibinfo {author} {\bibfnamefont {Wei-Yue}\ \bibnamefont {Liu}},
  \bibinfo {author} {\bibfnamefont {Liang}\ \bibnamefont {Zhang}}, \bibinfo
  {author} {\bibfnamefont {Yang}\ \bibnamefont {Li}}, \bibinfo {author}
  {\bibfnamefont {Ji-Gang}\ \bibnamefont {Ren}}, \bibinfo {author}
  {\bibfnamefont {Juan}\ \bibnamefont {Yin}}, \bibinfo {author} {\bibfnamefont
  {Qi}~\bibnamefont {Shen}}, \bibinfo {author} {\bibfnamefont {Yuan}\
  \bibnamefont {Cao}}, \bibinfo {author} {\bibfnamefont {Zheng-Ping}\
  \bibnamefont {Li}}, \bibinfo {author} {\bibfnamefont {Feng-Zhi}\ \bibnamefont
  {Li}}, \bibinfo {author} {\bibfnamefont {Xia-Wei}\ \bibnamefont {Chen}},
  \bibinfo {author} {\bibfnamefont {Li-Hua}\ \bibnamefont {Sun}}, \bibinfo
  {author} {\bibfnamefont {Jian-Jun}\ \bibnamefont {Jia}}, \bibinfo {author}
  {\bibfnamefont {Jin-Cai}\ \bibnamefont {Wu}}, \bibinfo {author}
  {\bibfnamefont {Xiao-Jun}\ \bibnamefont {Jiang}}, \bibinfo {author}
  {\bibfnamefont {Jian-Feng}\ \bibnamefont {Wang}}, \bibinfo {author}
  {\bibfnamefont {Yong-Mei}\ \bibnamefont {Huang}}, \bibinfo {author}
  {\bibfnamefont {Qiang}\ \bibnamefont {Wang}}, \bibinfo {author}
  {\bibfnamefont {Yi-Lin}\ \bibnamefont {Zhou}}, \bibinfo {author}
  {\bibfnamefont {Lei}\ \bibnamefont {Deng}}, \bibinfo {author} {\bibfnamefont
  {Tao}\ \bibnamefont {Xi}}, \bibinfo {author} {\bibfnamefont {Lu}~\bibnamefont
  {Ma}}, \bibinfo {author} {\bibfnamefont {Tai}\ \bibnamefont {Hu}}, \bibinfo
  {author} {\bibfnamefont {Qiang}\ \bibnamefont {Zhang}}, \bibinfo {author}
  {\bibfnamefont {Yu-Ao}\ \bibnamefont {Chen}}, \bibinfo {author}
  {\bibfnamefont {Nai-Le}\ \bibnamefont {Liu}}, \bibinfo {author}
  {\bibfnamefont {Xiang-Bin}\ \bibnamefont {Wang}}, \bibinfo {author}
  {\bibfnamefont {Zhen-Cai}\ \bibnamefont {Zhu}}, \bibinfo {author}
  {\bibfnamefont {Chao-Yang}\ \bibnamefont {Lu}}, \bibinfo {author}
  {\bibfnamefont {Rong}\ \bibnamefont {Shu}}, \bibinfo {author} {\bibfnamefont
  {Cheng-Zhi}\ \bibnamefont {Peng}}, \bibinfo {author} {\bibfnamefont
  {Jian-Yu}\ \bibnamefont {Wang}}, \ and\ \bibinfo {author} {\bibfnamefont
  {Jian-Wei}\ \bibnamefont {Pan}},\ }\bibfield  {title} {\enquote {\bibinfo
  {title} {Satellite-to-ground quantum key distribution},}\ }\href {\doibase
  10.1038/nature23655} {\bibfield  {journal} {\bibinfo  {journal} {Nature}\
  }\textbf {\bibinfo {volume} {549}},\ \bibinfo {pages} {43--47} (\bibinfo
  {year} {2017})},\ \bibinfo {note} {arXiv:1707.00542}\BibitemShut {NoStop}%
\bibitem [{\citenamefont {Tomamichel}\ \emph {et~al.}(2016)\citenamefont
  {Tomamichel}, \citenamefont {Berta},\ and\ \citenamefont {Renes}}]{TBR15}%
  \BibitemOpen
  \bibfield  {author} {\bibinfo {author} {\bibfnamefont {Marco}\ \bibnamefont
  {Tomamichel}}, \bibinfo {author} {\bibfnamefont {Mario}\ \bibnamefont
  {Berta}}, \ and\ \bibinfo {author} {\bibfnamefont {Joseph~M.}\ \bibnamefont
  {Renes}},\ }\bibfield  {title} {\enquote {\bibinfo {title} {Quantum coding
  with finite resources},}\ }\href {\doibase 10.1038/ncomms11419} {\bibfield
  {journal} {\bibinfo  {journal} {Nature Communications}\ }\textbf {\bibinfo
  {volume} {7}},\ \bibinfo {pages} {11419} (\bibinfo {year} {2016})},\ \bibinfo
  {note} {arXiv:1504.04617}\BibitemShut {NoStop}%
\bibitem [{\citenamefont {Wang}\ \emph {et~al.}(2019)\citenamefont {Wang},
  \citenamefont {Fang},\ and\ \citenamefont {Duan}}]{WFD17}%
  \BibitemOpen
  \bibfield  {author} {\bibinfo {author} {\bibfnamefont {Xin}\ \bibnamefont
  {Wang}}, \bibinfo {author} {\bibfnamefont {Kun}\ \bibnamefont {Fang}}, \ and\
  \bibinfo {author} {\bibfnamefont {Runyao}\ \bibnamefont {Duan}},\ }\bibfield
  {title} {\enquote {\bibinfo {title} {Semidefinite programming converse bounds
  for quantum communication},}\ }\href {\doibase 10.1109/TIT.2018.2874031}
  {\bibfield  {journal} {\bibinfo  {journal} {to appear in IEEE Transactions on
  Information Theory}\ } (\bibinfo {year} {2019}),\ 10.1109/TIT.2018.2874031},\
  \bibinfo {note} {arXiv:1709.00200}\BibitemShut {NoStop}%
\bibitem [{\citenamefont {Bennett}\ \emph {et~al.}(1996)\citenamefont
  {Bennett}, \citenamefont {DiVincenzo}, \citenamefont {Smolin},\ and\
  \citenamefont {Wootters}}]{BDSW96}%
  \BibitemOpen
  \bibfield  {author} {\bibinfo {author} {\bibfnamefont {Charles~H.}\
  \bibnamefont {Bennett}}, \bibinfo {author} {\bibfnamefont {David~P.}\
  \bibnamefont {DiVincenzo}}, \bibinfo {author} {\bibfnamefont {John~A.}\
  \bibnamefont {Smolin}}, \ and\ \bibinfo {author} {\bibfnamefont {William~K.}\
  \bibnamefont {Wootters}},\ }\bibfield  {title} {\enquote {\bibinfo {title}
  {Mixed-state entanglement and quantum error correction},}\ }\href {\doibase
  10.1103/PhysRevA.54.3824} {\bibfield  {journal} {\bibinfo  {journal}
  {Physical Review A}\ }\textbf {\bibinfo {volume} {54}},\ \bibinfo {pages}
  {3824--3851} (\bibinfo {year} {1996})},\ \bibinfo {note}
  {arXiv:quant-ph/9604024}\BibitemShut {NoStop}%
\bibitem [{\citenamefont {Horodecki}\ \emph {et~al.}(2009)\citenamefont
  {Horodecki}, \citenamefont {Horodecki}, \citenamefont {Horodecki},\ and\
  \citenamefont {Horodecki}}]{HHHH09}%
  \BibitemOpen
  \bibfield  {author} {\bibinfo {author} {\bibfnamefont {Ryszard}\ \bibnamefont
  {Horodecki}}, \bibinfo {author} {\bibfnamefont {Pawe\l{}}\ \bibnamefont
  {Horodecki}}, \bibinfo {author} {\bibfnamefont {Micha\l{}}\ \bibnamefont
  {Horodecki}}, \ and\ \bibinfo {author} {\bibfnamefont {Karol}\ \bibnamefont
  {Horodecki}},\ }\bibfield  {title} {\enquote {\bibinfo {title} {Quantum
  entanglement},}\ }\href {\doibase 10.1103/RevModPhys.81.865} {\bibfield
  {journal} {\bibinfo  {journal} {Review of Modern Physics}\ }\textbf {\bibinfo
  {volume} {81}},\ \bibinfo {pages} {865--942} (\bibinfo {year} {2009})},\
  \bibinfo {note} {arXiv:quant-ph/0702225}\BibitemShut {NoStop}%
\bibitem [{\citenamefont {Rains}(1999)}]{Rai99}%
  \BibitemOpen
  \bibfield  {author} {\bibinfo {author} {\bibfnamefont {Eric~M.}\ \bibnamefont
  {Rains}},\ }\bibfield  {title} {\enquote {\bibinfo {title} {Bound on
  distillable entanglement},}\ }\href {\doibase 10.1103/PhysRevA.60.179}
  {\bibfield  {journal} {\bibinfo  {journal} {Physical Review A}\ }\textbf
  {\bibinfo {volume} {60}},\ \bibinfo {pages} {179--184} (\bibinfo {year}
  {1999})},\ \bibinfo {note} {arXiv:quant-ph/9809082}\BibitemShut {NoStop}%
\bibitem [{\citenamefont {Rains}(2001)}]{Rai01}%
  \BibitemOpen
  \bibfield  {author} {\bibinfo {author} {\bibfnamefont {Eric~M.}\ \bibnamefont
  {Rains}},\ }\bibfield  {title} {\enquote {\bibinfo {title} {A semidefinite
  program for distillable entanglement},}\ }\href {\doibase 10.1109/18.959270}
  {\bibfield  {journal} {\bibinfo  {journal} {IEEE Transactions on Information
  Theory}\ }\textbf {\bibinfo {volume} {47}},\ \bibinfo {pages} {2921--2933}
  (\bibinfo {year} {2001})},\ \bibinfo {note}
  {arXiv:quant-ph/0008047}\BibitemShut {NoStop}%
\bibitem [{\citenamefont {Chitambar}\ \emph {et~al.}(2014)\citenamefont
  {Chitambar}, \citenamefont {Leung}, \citenamefont {Man{\v{c}}inska},
  \citenamefont {Ozols},\ and\ \citenamefont {Winter}}]{CLM+14}%
  \BibitemOpen
  \bibfield  {author} {\bibinfo {author} {\bibfnamefont {Eric}\ \bibnamefont
  {Chitambar}}, \bibinfo {author} {\bibfnamefont {Debbie}\ \bibnamefont
  {Leung}}, \bibinfo {author} {\bibfnamefont {Laura}\ \bibnamefont
  {Man{\v{c}}inska}}, \bibinfo {author} {\bibfnamefont {Maris}\ \bibnamefont
  {Ozols}}, \ and\ \bibinfo {author} {\bibfnamefont {Andreas}\ \bibnamefont
  {Winter}},\ }\bibfield  {title} {\enquote {\bibinfo {title} {Everything you
  always wanted to know about {LOCC} (but were afraid to ask)},}\ }\href
  {\doibase 10.1007/s00220-014-1953-9} {\bibfield  {journal} {\bibinfo
  {journal} {Communications in Mathematical Physics}\ }\textbf {\bibinfo
  {volume} {328}},\ \bibinfo {pages} {303--326} (\bibinfo {year} {2014})},\
  \bibinfo {note} {arXiv:1210.4583}\BibitemShut {NoStop}%
\bibitem [{\citenamefont {Kaur}\ \emph {et~al.}(2018)\citenamefont {Kaur},
  \citenamefont {Das}, \citenamefont {Wilde},\ and\ \citenamefont
  {Winter}}]{KDWW18}%
  \BibitemOpen
  \bibfield  {author} {\bibinfo {author} {\bibfnamefont {Eneet}\ \bibnamefont
  {Kaur}}, \bibinfo {author} {\bibfnamefont {Siddhartha}\ \bibnamefont {Das}},
  \bibinfo {author} {\bibfnamefont {Mark~M.}\ \bibnamefont {Wilde}}, \ and\
  \bibinfo {author} {\bibfnamefont {Andreas}\ \bibnamefont {Winter}},\
  }\href@noop {} {\  (\bibinfo {year} {2018})},\ \bibinfo {note}
  {arXiv:1803.10710}\BibitemShut {NoStop}%
\bibitem [{\citenamefont {Werner}(1989)}]{W89a}%
  \BibitemOpen
  \bibfield  {author} {\bibinfo {author} {\bibfnamefont {Reinhard~F.}\
  \bibnamefont {Werner}},\ }\bibfield  {title} {\enquote {\bibinfo {title} {An
  application of {Bell's} inequalities to a quantum state extension problem},}\
  }\href {\doibase 10.1007/BF00399761} {\bibfield  {journal} {\bibinfo
  {journal} {Letters in Mathematical Physics}\ }\textbf {\bibinfo {volume}
  {17}},\ \bibinfo {pages} {359--363} (\bibinfo {year} {1989})}\BibitemShut
  {NoStop}%
\bibitem [{\citenamefont {Doherty}\ \emph {et~al.}(2002)\citenamefont
  {Doherty}, \citenamefont {Parrilo},\ and\ \citenamefont
  {Spedalieri}}]{DPS02}%
  \BibitemOpen
  \bibfield  {author} {\bibinfo {author} {\bibfnamefont {Andrew~C.}\
  \bibnamefont {Doherty}}, \bibinfo {author} {\bibfnamefont {Pablo~A.}\
  \bibnamefont {Parrilo}}, \ and\ \bibinfo {author} {\bibfnamefont
  {Federico~M.}\ \bibnamefont {Spedalieri}},\ }\bibfield  {title} {\enquote
  {\bibinfo {title} {Distinguishing separable and entangled states},}\ }\href
  {\doibase 10.1103/PhysRevLett.88.187904} {\bibfield  {journal} {\bibinfo
  {journal} {Physical Review Letters}\ }\textbf {\bibinfo {volume} {88}},\
  \bibinfo {pages} {187904} (\bibinfo {year} {2002})},\ \bibinfo {note}
  {arXiv:quant-ph/0112007}\BibitemShut {NoStop}%
\bibitem [{\citenamefont {Doherty}\ \emph {et~al.}(2004)\citenamefont
  {Doherty}, \citenamefont {Parrilo},\ and\ \citenamefont
  {Spedalieri}}]{DPS04}%
  \BibitemOpen
  \bibfield  {author} {\bibinfo {author} {\bibfnamefont {Andrew~C.}\
  \bibnamefont {Doherty}}, \bibinfo {author} {\bibfnamefont {Pablo~A.}\
  \bibnamefont {Parrilo}}, \ and\ \bibinfo {author} {\bibfnamefont
  {Federico~M.}\ \bibnamefont {Spedalieri}},\ }\bibfield  {title} {\enquote
  {\bibinfo {title} {Complete family of separability criteria},}\ }\href
  {\doibase 10.1103/PhysRevA.69.022308} {\bibfield  {journal} {\bibinfo
  {journal} {Physical Review A}\ }\textbf {\bibinfo {volume} {69}},\ \bibinfo
  {pages} {022308} (\bibinfo {year} {2004})},\ \bibinfo {note}
  {arXiv:quant-ph/0308032}\BibitemShut {NoStop}%
\bibitem [{\citenamefont {van Loock}\ and\ \citenamefont
  {Braunstein}(2000)}]{vLB00}%
  \BibitemOpen
  \bibfield  {author} {\bibinfo {author} {\bibfnamefont {Peter}\ \bibnamefont
  {van Loock}}\ and\ \bibinfo {author} {\bibfnamefont {Samuel~L.}\ \bibnamefont
  {Braunstein}},\ }\bibfield  {title} {\enquote {\bibinfo {title} {Multipartite
  entanglement for continuous variables: A quantum teleportation network},}\
  }\href {\doibase 10.1103/PhysRevLett.84.3482} {\bibfield  {journal} {\bibinfo
   {journal} {Physical Review Letters}\ }\textbf {\bibinfo {volume} {84}},\
  \bibinfo {pages} {3482--3485} (\bibinfo {year} {2000})},\ \bibinfo {note}
  {arXiv:quant-ph/9906021}\BibitemShut {NoStop}%
\bibitem [{\citenamefont {Scarani}\ \emph {et~al.}(2005)\citenamefont
  {Scarani}, \citenamefont {Iblisdir}, \citenamefont {Gisin},\ and\
  \citenamefont {Ac\'{\i}n}}]{RevModPhys.77.1225}%
  \BibitemOpen
  \bibfield  {author} {\bibinfo {author} {\bibfnamefont {Valerio}\ \bibnamefont
  {Scarani}}, \bibinfo {author} {\bibfnamefont {Sofyan}\ \bibnamefont
  {Iblisdir}}, \bibinfo {author} {\bibfnamefont {Nicolas}\ \bibnamefont
  {Gisin}}, \ and\ \bibinfo {author} {\bibfnamefont {Antonio}\ \bibnamefont
  {Ac\'{\i}n}},\ }\bibfield  {title} {\enquote {\bibinfo {title} {Quantum
  cloning},}\ }\href {\doibase 10.1103/RevModPhys.77.1225} {\bibfield
  {journal} {\bibinfo  {journal} {Reviews of Modern Physics}\ }\textbf
  {\bibinfo {volume} {77}},\ \bibinfo {pages} {1225--1256} (\bibinfo {year}
  {2005})},\ \bibinfo {note} {arXiv:quant-ph/0511088}\BibitemShut {NoStop}%
\bibitem [{\citenamefont {Grassl}\ \emph {et~al.}(1997)\citenamefont {Grassl},
  \citenamefont {Beth},\ and\ \citenamefont {Pellizzari}}]{GBP97}%
  \BibitemOpen
  \bibfield  {author} {\bibinfo {author} {\bibfnamefont {Markus}\ \bibnamefont
  {Grassl}}, \bibinfo {author} {\bibfnamefont {Thomas}\ \bibnamefont {Beth}}, \
  and\ \bibinfo {author} {\bibfnamefont {Thomas}\ \bibnamefont {Pellizzari}},\
  }\bibfield  {title} {\enquote {\bibinfo {title} {Codes for the quantum
  erasure channel},}\ }\href {\doibase 10.1103/PhysRevA.56.33} {\bibfield
  {journal} {\bibinfo  {journal} {Physical Review A}\ }\textbf {\bibinfo
  {volume} {56}},\ \bibinfo {pages} {33} (\bibinfo {year} {1997})},\ \bibinfo
  {note} {arXiv:quant-ph/9610042}\BibitemShut {NoStop}%
\bibitem [{\citenamefont {Bennett}\ \emph {et~al.}(1997)\citenamefont
  {Bennett}, \citenamefont {DiVincenzo},\ and\ \citenamefont
  {Smolin}}]{PhysRevLett.78.3217}%
  \BibitemOpen
  \bibfield  {author} {\bibinfo {author} {\bibfnamefont {Charles~H.}\
  \bibnamefont {Bennett}}, \bibinfo {author} {\bibfnamefont {David~P.}\
  \bibnamefont {DiVincenzo}}, \ and\ \bibinfo {author} {\bibfnamefont
  {John~A.}\ \bibnamefont {Smolin}},\ }\bibfield  {title} {\enquote {\bibinfo
  {title} {Capacities of quantum erasure channels},}\ }\href {\doibase
  10.1103/PhysRevLett.78.3217} {\bibfield  {journal} {\bibinfo  {journal}
  {Physical Review Letters}\ }\textbf {\bibinfo {volume} {78}},\ \bibinfo
  {pages} {3217--3220} (\bibinfo {year} {1997})},\ \bibinfo {note}
  {arXiv:quant-ph/9701015}\BibitemShut {NoStop}%
\bibitem [{\citenamefont {Nowakowski}\ and\ \citenamefont
  {Horodecki}(2009)}]{NH09}%
  \BibitemOpen
  \bibfield  {author} {\bibinfo {author} {\bibfnamefont {Marcin~L.}\
  \bibnamefont {Nowakowski}}\ and\ \bibinfo {author} {\bibfnamefont {Pawel}\
  \bibnamefont {Horodecki}},\ }\bibfield  {title} {\enquote {\bibinfo {title}
  {A simple test for quantum channel capacity},}\ }\href
  {http://stacks.iop.org/1751-8121/42/i=13/a=135306} {\bibfield  {journal}
  {\bibinfo  {journal} {Journal of Physics A: Mathematical and Theoretical}\
  }\textbf {\bibinfo {volume} {42}},\ \bibinfo {pages} {135306} (\bibinfo
  {year} {2009})},\ \bibinfo {note} {arXiv:quant-ph/0503070}\BibitemShut
  {NoStop}%
\bibitem [{\citenamefont {{G.~S.~L.~}Brand{\~{a}}o}\ and\ \citenamefont
  {Gour}(2015)}]{BG15}%
  \BibitemOpen
  \bibfield  {author} {\bibinfo {author} {\bibfnamefont {Fernando}\
  \bibnamefont {{G.~S.~L.~}Brand{\~{a}}o}}\ and\ \bibinfo {author}
  {\bibfnamefont {Gilad}\ \bibnamefont {Gour}},\ }\bibfield  {title} {\enquote
  {\bibinfo {title} {Reversible framework for quantum resource theories},}\
  }\href {\doibase 10.1103/physrevlett.115.070503} {\bibfield  {journal}
  {\bibinfo  {journal} {Physical Review Letters}\ }\textbf {\bibinfo {volume}
  {115}},\ \bibinfo {pages} {070503} (\bibinfo {year} {2015})},\ \bibinfo
  {note} {arXiv:1502.03149}\BibitemShut {NoStop}%
\bibitem [{\citenamefont {Polyanskiy}\ and\ \citenamefont
  {Verd\'u}(2010)}]{PV10}%
  \BibitemOpen
  \bibfield  {author} {\bibinfo {author} {\bibfnamefont {Yury}\ \bibnamefont
  {Polyanskiy}}\ and\ \bibinfo {author} {\bibfnamefont {Sergio}\ \bibnamefont
  {Verd\'u}},\ }\bibfield  {title} {\enquote {\bibinfo {title} {Arimoto channel
  coding converse and {R\'enyi} divergence},}\ }in\ \href {\doibase
  10.1109/ALLERTON.2010.5707067} {\emph {\bibinfo {booktitle} {Proceedings of
  the 48th Annual Allerton Conference on Communication, Control, and
  Computation}}}\ (\bibinfo {year} {2010})\ pp.\ \bibinfo {pages}
  {1327--1333}\BibitemShut {NoStop}%
\bibitem [{\citenamefont {Sharma}\ and\ \citenamefont {Warsi}(2012)}]{SW12}%
  \BibitemOpen
  \bibfield  {author} {\bibinfo {author} {\bibfnamefont {Naresh}\ \bibnamefont
  {Sharma}}\ and\ \bibinfo {author} {\bibfnamefont {Naqueeb~Ahmad}\
  \bibnamefont {Warsi}},\ }\bibfield  {title} {\enquote {\bibinfo {title} {On
  the strong converses for the quantum channel capacity theorems},}\
  }\href@noop {} {\  (\bibinfo {year} {2012})},\ \bibinfo {note}
  {arXiv:1205.1712}\BibitemShut {NoStop}%
\bibitem [{\citenamefont {Brandao}(2008)}]{B08}%
  \BibitemOpen
  \bibfield  {author} {\bibinfo {author} {\bibfnamefont {Fernando G. S.~L.}\
  \bibnamefont {Brandao}},\ }\emph {\bibinfo {title} {Entanglement Theory and
  the Quantum Simulation of Many-Body Physics}},\ \href@noop {} {Ph.D.
  thesis},\ \bibinfo  {school} {Imperial College} (\bibinfo {year} {2008}),\
  \bibinfo {note} {arXiv:0810.0026}\BibitemShut {NoStop}%
\bibitem [{\citenamefont {Moroder}\ \emph {et~al.}(2006)\citenamefont
  {Moroder}, \citenamefont {Curty},\ and\ \citenamefont
  {L\"utkenhaus}}]{PhysRevA.74.052301}%
  \BibitemOpen
  \bibfield  {author} {\bibinfo {author} {\bibfnamefont {Tobias}\ \bibnamefont
  {Moroder}}, \bibinfo {author} {\bibfnamefont {Marcos}\ \bibnamefont {Curty}},
  \ and\ \bibinfo {author} {\bibfnamefont {Norbert}\ \bibnamefont
  {L\"utkenhaus}},\ }\bibfield  {title} {\enquote {\bibinfo {title} {One-way
  quantum key distribution: Simple upper bound on the secret key rate},}\
  }\href {\doibase 10.1103/PhysRevA.74.052301} {\bibfield  {journal} {\bibinfo
  {journal} {Physical Review A}\ }\textbf {\bibinfo {volume} {74}},\ \bibinfo
  {pages} {052301} (\bibinfo {year} {2006})},\ \bibinfo {note}
  {arXiv:quant-ph/0603270}\BibitemShut {NoStop}%
\bibitem [{\citenamefont {Hayden}\ \emph {et~al.}(2013)\citenamefont {Hayden},
  \citenamefont {Milner},\ and\ \citenamefont {Wilde}}]{HMW13}%
  \BibitemOpen
  \bibfield  {author} {\bibinfo {author} {\bibfnamefont {Patrick}\ \bibnamefont
  {Hayden}}, \bibinfo {author} {\bibfnamefont {Kevin}\ \bibnamefont {Milner}},
  \ and\ \bibinfo {author} {\bibfnamefont {Mark~M.}\ \bibnamefont {Wilde}},\
  }\bibfield  {title} {\enquote {\bibinfo {title} {Two-message quantum
  interactive proofs and the quantum separability problem},}\ }in\ \href
  {\doibase 10.1109/CCC.2013.24} {\emph {\bibinfo {booktitle} {Proceedings of
  the 28th IEEE Conference on Computational Complexity}}}\ (\bibinfo {address}
  {Palo Alto, California, USA},\ \bibinfo {year} {2013})\ pp.\ \bibinfo {pages}
  {156--167},\ \bibinfo {note} {arXiv:1211.6120}\BibitemShut {NoStop}%
\bibitem [{\citenamefont {Buscemi}\ and\ \citenamefont {Datta}(2010)}]{BD10}%
  \BibitemOpen
  \bibfield  {author} {\bibinfo {author} {\bibfnamefont {Francesco}\
  \bibnamefont {Buscemi}}\ and\ \bibinfo {author} {\bibfnamefont {Nilanjana}\
  \bibnamefont {Datta}},\ }\bibfield  {title} {\enquote {\bibinfo {title} {The
  quantum capacity of channels with arbitrarily correlated noise},}\ }\href
  {\doibase 10.1109/TIT.2009.2039166} {\bibfield  {journal} {\bibinfo
  {journal} {IEEE Transactions on Information Theory}\ }\textbf {\bibinfo
  {volume} {56}},\ \bibinfo {pages} {1447--1460} (\bibinfo {year} {2010})},\
  \bibinfo {note} {arXiv:0902.0158}\BibitemShut {NoStop}%
\bibitem [{\citenamefont {Wang}\ and\ \citenamefont {Renner}(2012)}]{WR12}%
  \BibitemOpen
  \bibfield  {author} {\bibinfo {author} {\bibfnamefont {Ligong}\ \bibnamefont
  {Wang}}\ and\ \bibinfo {author} {\bibfnamefont {Renato}\ \bibnamefont
  {Renner}},\ }\bibfield  {title} {\enquote {\bibinfo {title} {One-shot
  classical-quantum capacity and hypothesis testing},}\ }\href {\doibase
  10.1103/PhysRevLett.108.200501} {\bibfield  {journal} {\bibinfo  {journal}
  {Physical Review Letters}\ }\textbf {\bibinfo {volume} {108}},\ \bibinfo
  {pages} {200501} (\bibinfo {year} {2012})},\ \bibinfo {note}
  {arXiv:1007.5456}\BibitemShut {NoStop}%
\bibitem [{\citenamefont {Datta}(2009{\natexlab{a}})}]{D09}%
  \BibitemOpen
  \bibfield  {author} {\bibinfo {author} {\bibfnamefont {Nilanjana}\
  \bibnamefont {Datta}},\ }\bibfield  {title} {\enquote {\bibinfo {title} {Min-
  and max-relative entropies and a new entanglement monotone},}\ }\href
  {\doibase 10.1109/TIT.2009.2018325} {\bibfield  {journal} {\bibinfo
  {journal} {IEEE Transactions on Information Theory}\ }\textbf {\bibinfo
  {volume} {55}},\ \bibinfo {pages} {2816--2826} (\bibinfo {year}
  {2009}{\natexlab{a}})},\ \bibinfo {note} {arXiv:0803.2770}\BibitemShut
  {NoStop}%
\bibitem [{\citenamefont {Datta}(2009{\natexlab{b}})}]{Dat09}%
  \BibitemOpen
  \bibfield  {author} {\bibinfo {author} {\bibfnamefont {Nilanjana}\
  \bibnamefont {Datta}},\ }\bibfield  {title} {\enquote {\bibinfo {title}
  {Max-relative entropy of entanglement, alias log robustness},}\ }\href
  {\doibase 10.1142/S0219749909005298} {\bibfield  {journal} {\bibinfo
  {journal} {International Journal of Quantum Information}\ }\textbf {\bibinfo
  {volume} {7}},\ \bibinfo {pages} {475--491} (\bibinfo {year}
  {2009}{\natexlab{b}})},\ \bibinfo {note} {arXiv:0807.2536}\BibitemShut
  {NoStop}%
\bibitem [{\citenamefont {Smith}\ \emph {et~al.}(2008)\citenamefont {Smith},
  \citenamefont {Smolin},\ and\ \citenamefont {Winter}}]{SSW08}%
  \BibitemOpen
  \bibfield  {author} {\bibinfo {author} {\bibfnamefont {Graeme}\ \bibnamefont
  {Smith}}, \bibinfo {author} {\bibfnamefont {John~A.}\ \bibnamefont {Smolin}},
  \ and\ \bibinfo {author} {\bibfnamefont {Andreas}\ \bibnamefont {Winter}},\
  }\bibfield  {title} {\enquote {\bibinfo {title} {The quantum capacity with
  symmetric side channels},}\ }\href {\doibase 10.1109/TIT.2008.928269}
  {\bibfield  {journal} {\bibinfo  {journal} {IEEE Transactions on Information
  Theory}\ }\textbf {\bibinfo {volume} {54}},\ \bibinfo {pages} {4208--4217}
  (\bibinfo {year} {2008})},\ \bibinfo {note}
  {arXiv:quant-ph/0607039}\BibitemShut {NoStop}%
\bibitem [{\citenamefont {Schumacher}(1996)}]{PhysRevA.54.2614}%
  \BibitemOpen
  \bibfield  {author} {\bibinfo {author} {\bibfnamefont {Benjamin}\
  \bibnamefont {Schumacher}},\ }\bibfield  {title} {\enquote {\bibinfo {title}
  {Sending entanglement through noisy quantum channels},}\ }\href {\doibase
  10.1103/PhysRevA.54.2614} {\bibfield  {journal} {\bibinfo  {journal}
  {Physical Review A}\ }\textbf {\bibinfo {volume} {54}},\ \bibinfo {pages}
  {2614--2628} (\bibinfo {year} {1996})}\BibitemShut {NoStop}%
\bibitem [{\citenamefont {Schumacher}\ and\ \citenamefont
  {Nielsen}(1996)}]{PhysRevA.54.2629}%
  \BibitemOpen
  \bibfield  {author} {\bibinfo {author} {\bibfnamefont {Benjamin}\
  \bibnamefont {Schumacher}}\ and\ \bibinfo {author} {\bibfnamefont
  {Michael~A.}\ \bibnamefont {Nielsen}},\ }\bibfield  {title} {\enquote
  {\bibinfo {title} {Quantum data processing and error correction},}\ }\href
  {\doibase 10.1103/PhysRevA.54.2629} {\bibfield  {journal} {\bibinfo
  {journal} {Physical Review A}\ }\textbf {\bibinfo {volume} {54}},\ \bibinfo
  {pages} {2629--2635} (\bibinfo {year} {1996})},\ \bibinfo {note}
  {arXiv:quant-ph/9604022}\BibitemShut {NoStop}%
\bibitem [{\citenamefont {Lloyd}(1997)}]{L97}%
  \BibitemOpen
  \bibfield  {author} {\bibinfo {author} {\bibfnamefont {Seth}\ \bibnamefont
  {Lloyd}},\ }\bibfield  {title} {\enquote {\bibinfo {title} {Capacity of the
  noisy quantum channel},}\ }\href {\doibase 10.1103/PhysRevA.55.1613}
  {\bibfield  {journal} {\bibinfo  {journal} {Physical Review A}\ }\textbf
  {\bibinfo {volume} {55}},\ \bibinfo {pages} {1613} (\bibinfo {year}
  {1997})},\ \bibinfo {note} {arXiv:quant-ph/9604015}\BibitemShut {NoStop}%
\bibitem [{\citenamefont {Barnum}\ \emph {et~al.}(1998)\citenamefont {Barnum},
  \citenamefont {Nielsen},\ and\ \citenamefont {Schumacher}}]{BNS98}%
  \BibitemOpen
  \bibfield  {author} {\bibinfo {author} {\bibfnamefont {Howard}\ \bibnamefont
  {Barnum}}, \bibinfo {author} {\bibfnamefont {M.~A.}\ \bibnamefont {Nielsen}},
  \ and\ \bibinfo {author} {\bibfnamefont {Benjamin}\ \bibnamefont
  {Schumacher}},\ }\bibfield  {title} {\enquote {\bibinfo {title} {Information
  transmission through a noisy quantum channel},}\ }\href {\doibase
  10.1103/PhysRevA.57.4153} {\bibfield  {journal} {\bibinfo  {journal}
  {Physical Review A}\ }\textbf {\bibinfo {volume} {57}},\ \bibinfo {pages}
  {4153--4175} (\bibinfo {year} {1998})}\BibitemShut {NoStop}%
\bibitem [{\citenamefont {Barnum}\ \emph {et~al.}(2000)\citenamefont {Barnum},
  \citenamefont {Knill},\ and\ \citenamefont {Nielsen}}]{BKN98}%
  \BibitemOpen
  \bibfield  {author} {\bibinfo {author} {\bibfnamefont {Howard}\ \bibnamefont
  {Barnum}}, \bibinfo {author} {\bibfnamefont {Emanuel}\ \bibnamefont {Knill}},
  \ and\ \bibinfo {author} {\bibfnamefont {Michael~A.}\ \bibnamefont
  {Nielsen}},\ }\bibfield  {title} {\enquote {\bibinfo {title} {On quantum
  fidelities and channel capacities},}\ }\href {\doibase 10.1109/18.850671}
  {\bibfield  {journal} {\bibinfo  {journal} {IEEE Transactions on Information
  Theory}\ }\textbf {\bibinfo {volume} {46}},\ \bibinfo {pages} {1317}
  (\bibinfo {year} {2000})},\ \bibinfo {note}
  {arXiv:quant-ph/9809010}\BibitemShut {NoStop}%
\bibitem [{\citenamefont {Shor}(2002)}]{Sho02}%
  \BibitemOpen
  \bibfield  {author} {\bibinfo {author} {\bibfnamefont {Peter~W.}\
  \bibnamefont {Shor}},\ }\bibfield  {title} {\enquote {\bibinfo {title} {The
  quantum channel capacity and coherent information},}\ }in\ \href@noop {}
  {\emph {\bibinfo {booktitle} {Lecture Notes, MSRI Workshop on Quantum
  Computation}}}\ (\bibinfo {year} {2002})\BibitemShut {NoStop}%
\bibitem [{\citenamefont {Devetak}(2005)}]{D05}%
  \BibitemOpen
  \bibfield  {author} {\bibinfo {author} {\bibfnamefont {Igor}\ \bibnamefont
  {Devetak}},\ }\bibfield  {title} {\enquote {\bibinfo {title} {The private
  classical capacity and quantum capacity of a quantum channel},}\ }\href
  {\doibase 10.1109/TIT.2004.839515} {\bibfield  {journal} {\bibinfo  {journal}
  {IEEE Transactions on Information Theory}\ }\textbf {\bibinfo {volume}
  {51}},\ \bibinfo {pages} {44--55} (\bibinfo {year} {2005})},\ \bibinfo {note}
  {arXiv:quant-ph/0304127}\BibitemShut {NoStop}%
\bibitem [{\citenamefont {Uhlmann}(1976)}]{U76}%
  \BibitemOpen
  \bibfield  {author} {\bibinfo {author} {\bibfnamefont {Armin}\ \bibnamefont
  {Uhlmann}},\ }\bibfield  {title} {\enquote {\bibinfo {title} {The
  ``transition probability'' in the state space of a *-algebra},}\ }\href
  {\doibase 10.1016/0034-4877(76)90060-4} {\bibfield  {journal} {\bibinfo
  {journal} {Reports on Mathematical Physics}\ }\textbf {\bibinfo {volume}
  {9}},\ \bibinfo {pages} {273--279} (\bibinfo {year} {1976})}\BibitemShut
  {NoStop}%
\bibitem [{\citenamefont {Takeoka}\ \emph
  {et~al.}(2014{\natexlab{a}})\citenamefont {Takeoka}, \citenamefont {Guha},\
  and\ \citenamefont {Wilde}}]{TGW14}%
  \BibitemOpen
  \bibfield  {author} {\bibinfo {author} {\bibfnamefont {Masahiro}\
  \bibnamefont {Takeoka}}, \bibinfo {author} {\bibfnamefont {Saikat}\
  \bibnamefont {Guha}}, \ and\ \bibinfo {author} {\bibfnamefont {Mark~M.}\
  \bibnamefont {Wilde}},\ }\bibfield  {title} {\enquote {\bibinfo {title} {The
  squashed entanglement of a quantum channel},}\ }\href {\doibase
  10.1109/TIT.2014.2330313} {\bibfield  {journal} {\bibinfo  {journal} {IEEE
  Transactions on Information Theory}\ }\textbf {\bibinfo {volume} {60}},\
  \bibinfo {pages} {4987--4998} (\bibinfo {year} {2014}{\natexlab{a}})},\
  \bibinfo {note} {arXiv:1310.0129}\BibitemShut {NoStop}%
\bibitem [{\citenamefont {Takeoka}\ \emph
  {et~al.}(2014{\natexlab{b}})\citenamefont {Takeoka}, \citenamefont {Guha},\
  and\ \citenamefont {Wilde}}]{TGW14nat}%
  \BibitemOpen
  \bibfield  {author} {\bibinfo {author} {\bibfnamefont {Masahiro}\
  \bibnamefont {Takeoka}}, \bibinfo {author} {\bibfnamefont {Saikat}\
  \bibnamefont {Guha}}, \ and\ \bibinfo {author} {\bibfnamefont {Mark~M.}\
  \bibnamefont {Wilde}},\ }\bibfield  {title} {\enquote {\bibinfo {title}
  {Fundamental rate-loss tradeoff for optical quantum key distribution},}\
  }\href {\doibase 10.1038/ncomms6235} {\bibfield  {journal} {\bibinfo
  {journal} {Nature Communications}\ }\textbf {\bibinfo {volume} {5}},\
  \bibinfo {pages} {5235} (\bibinfo {year} {2014}{\natexlab{b}})},\ \bibinfo
  {note} {arXiv:1504.06390}\BibitemShut {NoStop}%
\bibitem [{\citenamefont {Kaur}\ and\ \citenamefont {Wilde}(2018)}]{KW17a}%
  \BibitemOpen
  \bibfield  {author} {\bibinfo {author} {\bibfnamefont {Eneet}\ \bibnamefont
  {Kaur}}\ and\ \bibinfo {author} {\bibfnamefont {Mark~M.}\ \bibnamefont
  {Wilde}},\ }\bibfield  {title} {\enquote {\bibinfo {title} {Amortized
  entanglement of a quantum channel and approximately teleportation-simulable
  channels},}\ }\href {\doibase 10.1088/1751-8121/aa9da7} {\bibfield  {journal}
  {\bibinfo  {journal} {Journal of Physics A}\ }\textbf {\bibinfo {volume}
  {51}},\ \bibinfo {pages} {035303} (\bibinfo {year} {2018})},\ \bibinfo {note}
  {arXiv:1707.07721}\BibitemShut {NoStop}%
\bibitem [{\citenamefont {Johnson}\ and\ \citenamefont {Viola}(2013)}]{JV13}%
  \BibitemOpen
  \bibfield  {author} {\bibinfo {author} {\bibfnamefont {Peter~D.}\
  \bibnamefont {Johnson}}\ and\ \bibinfo {author} {\bibfnamefont {Lorenza}\
  \bibnamefont {Viola}},\ }\bibfield  {title} {\enquote {\bibinfo {title}
  {Compatible quantum correlations: Extension problems for {Werner} and
  isotropic states},}\ }\href {\doibase 10.1103/physreva.88.032323} {\bibfield
  {journal} {\bibinfo  {journal} {Physical Review A}\ }\textbf {\bibinfo
  {volume} {88}},\ \bibinfo {pages} {032323} (\bibinfo {year} {2013})},\
  \bibinfo {note} {arXiv:1305.1342}\BibitemShut {NoStop}%
\bibitem [{\citenamefont {Morgan}\ and\ \citenamefont {Winter}(2014)}]{MW13}%
  \BibitemOpen
  \bibfield  {author} {\bibinfo {author} {\bibfnamefont {Ciara}\ \bibnamefont
  {Morgan}}\ and\ \bibinfo {author} {\bibfnamefont {Andreas}\ \bibnamefont
  {Winter}},\ }\bibfield  {title} {\enquote {\bibinfo {title} {{``Pretty
  strong''} converse for the quantum capacity of degradable channels},}\ }\href
  {\doibase 10.1109/TIT.2013.2288971} {\bibfield  {journal} {\bibinfo
  {journal} {IEEE Transactions on Information Theory}\ }\textbf {\bibinfo
  {volume} {60}},\ \bibinfo {pages} {317--333} (\bibinfo {year} {2014})},\
  \bibinfo {note} {arXiv:1301.4927}\BibitemShut {NoStop}%
\bibitem [{\citenamefont {Caruso}\ and\ \citenamefont
  {Giovannetti}(2006)}]{CG06}%
  \BibitemOpen
  \bibfield  {author} {\bibinfo {author} {\bibfnamefont {Filippo}\ \bibnamefont
  {Caruso}}\ and\ \bibinfo {author} {\bibfnamefont {Vittorio}\ \bibnamefont
  {Giovannetti}},\ }\bibfield  {title} {\enquote {\bibinfo {title}
  {Degradability of bosonic {Gaussian} channels},}\ }\href {\doibase
  10.1103/PhysRevA.74.062307} {\bibfield  {journal} {\bibinfo  {journal}
  {Physical Review A}\ }\textbf {\bibinfo {volume} {74}},\ \bibinfo {pages}
  {062307} (\bibinfo {year} {2006})},\ \bibinfo {note}
  {arXiv:quant-ph/0603257}\BibitemShut {NoStop}%
\bibitem [{\citenamefont {Myhr}(2010)}]{M10}%
  \BibitemOpen
  \bibfield  {author} {\bibinfo {author} {\bibfnamefont {Geir~Ove}\
  \bibnamefont {Myhr}},\ }\emph {\bibinfo {title} {Symmetric extension of
  bipartite quantum states and its use in quantum key distribution with two-way
  postprocessing}},\ \href@noop {} {Ph.D. thesis},\ \bibinfo  {school}
  {Universit\"at Erlangen-N\"urnberg} (\bibinfo {year} {2010}),\ \bibinfo
  {note} {arXiv:1103.0766}\BibitemShut {NoStop}%
\bibitem [{\citenamefont {Holevo}(2008)}]{Holevo2008}%
  \BibitemOpen
  \bibfield  {author} {\bibinfo {author} {\bibfnamefont {Alexander~S.}\
  \bibnamefont {Holevo}},\ }\bibfield  {title} {\enquote {\bibinfo {title}
  {Entanglement-breaking channels in infinite dimensions},}\ }\href {\doibase
  10.1134/S0032946008030010} {\bibfield  {journal} {\bibinfo  {journal}
  {Problems of Information Transmission}\ }\textbf {\bibinfo {volume} {44}},\
  \bibinfo {pages} {171--184} (\bibinfo {year} {2008})},\ \bibinfo {note}
  {arXiv:0802.0235}\BibitemShut {NoStop}%
\bibitem [{\citenamefont {Chow}\ \emph {et~al.}(2011)\citenamefont {Chow},
  \citenamefont {C\'orcoles}, \citenamefont {Gambetta}, \citenamefont
  {Rigetti}, \citenamefont {Johnson}, \citenamefont {Smolin}, \citenamefont
  {Rozen}, \citenamefont {Keefe}, \citenamefont {Rothwell}, \citenamefont
  {Ketchen},\ and\ \citenamefont {Steffen}}]{CCG+11}%
  \BibitemOpen
  \bibfield  {author} {\bibinfo {author} {\bibfnamefont {Jerry~M.}\
  \bibnamefont {Chow}}, \bibinfo {author} {\bibfnamefont {A.~D.}\ \bibnamefont
  {C\'orcoles}}, \bibinfo {author} {\bibfnamefont {Jay~M.}\ \bibnamefont
  {Gambetta}}, \bibinfo {author} {\bibfnamefont {Chad}\ \bibnamefont
  {Rigetti}}, \bibinfo {author} {\bibfnamefont {B.~R.}\ \bibnamefont
  {Johnson}}, \bibinfo {author} {\bibfnamefont {John~A.}\ \bibnamefont
  {Smolin}}, \bibinfo {author} {\bibfnamefont {J.~R.}\ \bibnamefont {Rozen}},
  \bibinfo {author} {\bibfnamefont {George~A.}\ \bibnamefont {Keefe}}, \bibinfo
  {author} {\bibfnamefont {Mary~B.}\ \bibnamefont {Rothwell}}, \bibinfo
  {author} {\bibfnamefont {Mark~B.}\ \bibnamefont {Ketchen}}, \ and\ \bibinfo
  {author} {\bibfnamefont {M.}~\bibnamefont {Steffen}},\ }\bibfield  {title}
  {\enquote {\bibinfo {title} {Simple all-microwave entangling gate for
  fixed-frequency superconducting qubits},}\ }\href {\doibase
  10.1103/PhysRevLett.107.080502} {\bibfield  {journal} {\bibinfo  {journal}
  {Physical Review Letters}\ }\textbf {\bibinfo {volume} {107}},\ \bibinfo
  {pages} {080502} (\bibinfo {year} {2011})},\ \bibinfo {note}
  {arXiv:1106.0553}\BibitemShut {NoStop}%
\bibitem [{\citenamefont {Linke}\ \emph {et~al.}(2017)\citenamefont {Linke},
  \citenamefont {Maslov}, \citenamefont {Roetteler}, \citenamefont {Debnath},
  \citenamefont {Figgatt}, \citenamefont {Landsman}, \citenamefont {Wright},\
  and\ \citenamefont {Monroe}}]{LMR+17}%
  \BibitemOpen
  \bibfield  {author} {\bibinfo {author} {\bibfnamefont {Norbert~M.}\
  \bibnamefont {Linke}}, \bibinfo {author} {\bibfnamefont {Dmitri}\
  \bibnamefont {Maslov}}, \bibinfo {author} {\bibfnamefont {Martin}\
  \bibnamefont {Roetteler}}, \bibinfo {author} {\bibfnamefont {Shantanu}\
  \bibnamefont {Debnath}}, \bibinfo {author} {\bibfnamefont {Caroline}\
  \bibnamefont {Figgatt}}, \bibinfo {author} {\bibfnamefont {Kevin~A.}\
  \bibnamefont {Landsman}}, \bibinfo {author} {\bibfnamefont {Kenneth}\
  \bibnamefont {Wright}}, \ and\ \bibinfo {author} {\bibfnamefont
  {Christopher}\ \bibnamefont {Monroe}},\ }\bibfield  {title} {\enquote
  {\bibinfo {title} {Experimental comparison of two quantum computing
  architectures},}\ }\href {\doibase 10.1073/pnas.1618020114} {\bibfield
  {journal} {\bibinfo  {journal} {Proceedings of the National Academy of
  Sciences}\ }\textbf {\bibinfo {volume} {114}},\ \bibinfo {pages} {3305--3310}
  (\bibinfo {year} {2017})},\ \bibinfo {note} {arXiv:1702.01852}\BibitemShut
  {NoStop}%
\bibitem [{\citenamefont {Barrett}\ and\ \citenamefont {Stace}(2010)}]{BS10}%
  \BibitemOpen
  \bibfield  {author} {\bibinfo {author} {\bibfnamefont {Sean~D.}\ \bibnamefont
  {Barrett}}\ and\ \bibinfo {author} {\bibfnamefont {Thomas~M.}\ \bibnamefont
  {Stace}},\ }\bibfield  {title} {\enquote {\bibinfo {title} {Fault tolerant
  quantum computation with very high threshold for loss errors},}\ }\href
  {\doibase 10.1103/PhysRevLett.105.200502} {\bibfield  {journal} {\bibinfo
  {journal} {Physical Review Letters}\ }\textbf {\bibinfo {volume} {105}},\
  \bibinfo {pages} {200502} (\bibinfo {year} {2010})},\ \bibinfo {note}
  {arXiv:1005.2456}\BibitemShut {NoStop}%
\bibitem [{\citenamefont {Dumer}\ \emph {et~al.}(2015)\citenamefont {Dumer},
  \citenamefont {Kovalev},\ and\ \citenamefont {Pryadko}}]{DKP15}%
  \BibitemOpen
  \bibfield  {author} {\bibinfo {author} {\bibfnamefont {Ilya}\ \bibnamefont
  {Dumer}}, \bibinfo {author} {\bibfnamefont {Alexey~A.}\ \bibnamefont
  {Kovalev}}, \ and\ \bibinfo {author} {\bibfnamefont {Leonid~P.}\ \bibnamefont
  {Pryadko}},\ }\bibfield  {title} {\enquote {\bibinfo {title} {Thresholds for
  correcting errors, erasures, and faulty syndrome measurements in degenerate
  quantum codes},}\ }\href {\doibase 10.1103/PhysRevLett.115.050502} {\bibfield
   {journal} {\bibinfo  {journal} {Physical Review Letters}\ }\textbf {\bibinfo
  {volume} {115}},\ \bibinfo {pages} {050502} (\bibinfo {year} {2015})},\
  \bibinfo {note} {arXiv:1412.6172}\BibitemShut {NoStop}%
\bibitem [{\citenamefont {{Holevo}}(2002)}]{H02}%
  \BibitemOpen
  \bibfield  {author} {\bibinfo {author} {\bibfnamefont {Alexander~S.}\
  \bibnamefont {{Holevo}}},\ }\bibfield  {title} {\enquote {\bibinfo {title}
  {Remarks on the classical capacity of quantum channel},}\ }\href@noop {} {\
  (\bibinfo {year} {2002})},\ \bibinfo {note} {arXiv:quant-ph/0212025},\
  \Eprint {http://arxiv.org/abs/quant-ph/0212025} {quant-ph/0212025}
  \BibitemShut {NoStop}%
\bibitem [{\citenamefont {Chiribella}\ \emph {et~al.}(2009)\citenamefont
  {Chiribella}, \citenamefont {D'Ariano},\ and\ \citenamefont
  {Perinotti}}]{CDP09}%
  \BibitemOpen
  \bibfield  {author} {\bibinfo {author} {\bibfnamefont {Giulio}\ \bibnamefont
  {Chiribella}}, \bibinfo {author} {\bibfnamefont {Giacomo~Mauro}\ \bibnamefont
  {D'Ariano}}, \ and\ \bibinfo {author} {\bibfnamefont {Paolo}\ \bibnamefont
  {Perinotti}},\ }\bibfield  {title} {\enquote {\bibinfo {title} {Realization
  schemes for quantum instruments in finite dimensions},}\ }\href {\doibase
  10.1063/1.3105923} {\bibfield  {journal} {\bibinfo  {journal} {Journal of
  Mathematical Physics}\ }\textbf {\bibinfo {volume} {50}},\ \bibinfo {pages}
  {042101} (\bibinfo {year} {2009})},\ \bibinfo {note} {arXiv:0810.3211},\
  \Eprint {http://arxiv.org/abs/http://dx.doi.org/10.1063/1.3105923}
  {http://dx.doi.org/10.1063/1.3105923} \BibitemShut {NoStop}%
\bibitem [{\citenamefont {Leditzky}\ \emph
  {et~al.}(2018{\natexlab{a}})\citenamefont {Leditzky}, \citenamefont {Datta},\
  and\ \citenamefont {Smith}}]{LDS17}%
  \BibitemOpen
  \bibfield  {author} {\bibinfo {author} {\bibfnamefont {Felix}\ \bibnamefont
  {Leditzky}}, \bibinfo {author} {\bibfnamefont {Nilanjana}\ \bibnamefont
  {Datta}}, \ and\ \bibinfo {author} {\bibfnamefont {Graeme}\ \bibnamefont
  {Smith}},\ }\bibfield  {title} {\enquote {\bibinfo {title} {Useful states and
  entanglement distillation},}\ }\href {\doibase 10.1109/TIT.2017.2776907}
  {\bibfield  {journal} {\bibinfo  {journal} {IEEE Transactions on Information
  Theory}\ } (\bibinfo {year} {2018}{\natexlab{a}}),\
  10.1109/TIT.2017.2776907},\ \bibinfo {note} {arXiv:1701.03081}\BibitemShut
  {NoStop}%
\bibitem [{\citenamefont {Leditzky}\ \emph
  {et~al.}(2018{\natexlab{b}})\citenamefont {Leditzky}, \citenamefont {Leung},\
  and\ \citenamefont {Smith}}]{LLS17}%
  \BibitemOpen
  \bibfield  {author} {\bibinfo {author} {\bibfnamefont {Felix}\ \bibnamefont
  {Leditzky}}, \bibinfo {author} {\bibfnamefont {Debbie}\ \bibnamefont
  {Leung}}, \ and\ \bibinfo {author} {\bibfnamefont {Graeme}\ \bibnamefont
  {Smith}},\ }\bibfield  {title} {\enquote {\bibinfo {title} {Quantum and
  private capacities of low-noise channels},}\ }\href {\doibase
  10.1103/PhysRevLett.120.160503} {\bibfield  {journal} {\bibinfo  {journal}
  {Physical Review Letters}\ }\textbf {\bibinfo {volume} {120}},\ \bibinfo
  {pages} {160503} (\bibinfo {year} {2018}{\natexlab{b}})},\ \bibinfo {note}
  {arXiv:1705.04335}\BibitemShut {NoStop}%
\bibitem [{\citenamefont {Bru\ss{}}\ \emph {et~al.}(1998)\citenamefont
  {Bru\ss{}}, \citenamefont {DiVincenzo}, \citenamefont {Ekert}, \citenamefont
  {Fuchs}, \citenamefont {Macchiavello},\ and\ \citenamefont
  {Smolin}}]{BDEFMS98}%
  \BibitemOpen
  \bibfield  {author} {\bibinfo {author} {\bibfnamefont {Dagmar}\ \bibnamefont
  {Bru\ss{}}}, \bibinfo {author} {\bibfnamefont {David~P.}\ \bibnamefont
  {DiVincenzo}}, \bibinfo {author} {\bibfnamefont {Artur}\ \bibnamefont
  {Ekert}}, \bibinfo {author} {\bibfnamefont {Christopher~A.}\ \bibnamefont
  {Fuchs}}, \bibinfo {author} {\bibfnamefont {Chiara}\ \bibnamefont
  {Macchiavello}}, \ and\ \bibinfo {author} {\bibfnamefont {John~A.}\
  \bibnamefont {Smolin}},\ }\bibfield  {title} {\enquote {\bibinfo {title}
  {Optimal universal and state-dependent quantum cloning},}\ }\href {\doibase
  10.1103/PhysRevA.57.2368} {\bibfield  {journal} {\bibinfo  {journal}
  {Physical Review A}\ }\textbf {\bibinfo {volume} {57}},\ \bibinfo {pages}
  {2368--2378} (\bibinfo {year} {1998})},\ \bibinfo {note}
  {arXiv:quant-ph/9705038}\BibitemShut {NoStop}%
\bibitem [{\citenamefont {Cerf}(2000)}]{C00}%
  \BibitemOpen
  \bibfield  {author} {\bibinfo {author} {\bibfnamefont {Nicolas~J.}\
  \bibnamefont {Cerf}},\ }\bibfield  {title} {\enquote {\bibinfo {title} {Pauli
  cloning of a quantum bit},}\ }\href {\doibase 10.1103/PhysRevLett.84.4497}
  {\bibfield  {journal} {\bibinfo  {journal} {Physical Review Letters}\
  }\textbf {\bibinfo {volume} {84}},\ \bibinfo {pages} {4497--4500} (\bibinfo
  {year} {2000})},\ \bibinfo {note} {arXiv:quant-ph/9803058}\BibitemShut
  {NoStop}%
\bibitem [{\citenamefont {Polyanskiy}\ \emph {et~al.}(2010)\citenamefont
  {Polyanskiy}, \citenamefont {Poor},\ and\ \citenamefont {Verd\'{u}}}]{PPV10}%
  \BibitemOpen
  \bibfield  {author} {\bibinfo {author} {\bibfnamefont {Yury}\ \bibnamefont
  {Polyanskiy}}, \bibinfo {author} {\bibfnamefont {H.~Vincent}\ \bibnamefont
  {Poor}}, \ and\ \bibinfo {author} {\bibfnamefont {Sergio}\ \bibnamefont
  {Verd\'{u}}},\ }\bibfield  {title} {\enquote {\bibinfo {title} {Channel
  coding rate in the finite blocklength regime},}\ }\href {\doibase
  10.1109/TIT.2010.2043769} {\bibfield  {journal} {\bibinfo  {journal} {IEEE
  Transactions on Information Theory}\ }\textbf {\bibinfo {volume} {56}},\
  \bibinfo {pages} {2307--2359} (\bibinfo {year} {2010})}\BibitemShut {NoStop}%
\bibitem [{\citenamefont {Matthews}\ and\ \citenamefont {Wehner}(2014)}]{MW14}%
  \BibitemOpen
  \bibfield  {author} {\bibinfo {author} {\bibfnamefont {William}\ \bibnamefont
  {Matthews}}\ and\ \bibinfo {author} {\bibfnamefont {Stephanie}\ \bibnamefont
  {Wehner}},\ }\bibfield  {title} {\enquote {\bibinfo {title} {Finite
  blocklength converse bounds for quantum channels},}\ }\href {\doibase
  10.1109/tit.2014.2353614} {\bibfield  {journal} {\bibinfo  {journal} {{IEEE}
  Transactions on Information Theory}\ }\textbf {\bibinfo {volume} {60}},\
  \bibinfo {pages} {7317--7329} (\bibinfo {year} {2014})},\ \bibinfo {note}
  {arXiv:1210.4722}\BibitemShut {NoStop}%
\bibitem [{\citenamefont {Wilde}\ \emph {et~al.}(2017)\citenamefont {Wilde},
  \citenamefont {Tomamichel},\ and\ \citenamefont {Berta}}]{WTB17}%
  \BibitemOpen
  \bibfield  {author} {\bibinfo {author} {\bibfnamefont {Mark~M.}\ \bibnamefont
  {Wilde}}, \bibinfo {author} {\bibfnamefont {Marco}\ \bibnamefont
  {Tomamichel}}, \ and\ \bibinfo {author} {\bibfnamefont {Mario}\ \bibnamefont
  {Berta}},\ }\bibfield  {title} {\enquote {\bibinfo {title} {Converse bounds
  for private communication over quantum channels},}\ }\href {\doibase
  10.1109/TIT.2017.2648825} {\bibfield  {journal} {\bibinfo  {journal} {IEEE
  Transactions on Information Theory}\ }\textbf {\bibinfo {volume} {63}},\
  \bibinfo {pages} {1792--1817} (\bibinfo {year} {2017})},\ \bibinfo {note}
  {arXiv:1602.08898}\BibitemShut {NoStop}%
\bibitem [{\citenamefont {Das}\ \emph {et~al.}(2017)\citenamefont {Das},
  \citenamefont {B{\"a}uml},\ and\ \citenamefont {Wilde}}]{DBW17}%
  \BibitemOpen
  \bibfield  {author} {\bibinfo {author} {\bibfnamefont {Siddhartha}\
  \bibnamefont {Das}}, \bibinfo {author} {\bibfnamefont {Stefan}\ \bibnamefont
  {B{\"a}uml}}, \ and\ \bibinfo {author} {\bibfnamefont {Mark~M.}\ \bibnamefont
  {Wilde}},\ }\bibfield  {title} {\enquote {\bibinfo {title} {Entanglement and
  secret-key-agreement capacities of bipartite quantum interactions and
  read-only memory devices},}\ }\href@noop {} {\  (\bibinfo {year} {2017})},\
  \bibinfo {note} {arXiv:1712.00827}\BibitemShut {NoStop}%
\bibitem [{\citenamefont {B{\"a}uml}\ \emph {et~al.}(2018)\citenamefont
  {B{\"a}uml}, \citenamefont {Das},\ and\ \citenamefont {Wilde}}]{BDW18}%
  \BibitemOpen
  \bibfield  {author} {\bibinfo {author} {\bibfnamefont {Stefan}\ \bibnamefont
  {B{\"a}uml}}, \bibinfo {author} {\bibfnamefont {Siddhartha}\ \bibnamefont
  {Das}}, \ and\ \bibinfo {author} {\bibfnamefont {Mark~M.}\ \bibnamefont
  {Wilde}},\ }\bibfield  {title} {\enquote {\bibinfo {title} {Fundamental
  limits on the capacities of bipartite quantum interactions},}\ }\href
  {\doibase 10.1103/PhysRevLett.121.250504} {\bibfield  {journal} {\bibinfo
  {journal} {Physical Review Letters}\ }\textbf {\bibinfo {volume} {121}},\
  \bibinfo {pages} {250504} (\bibinfo {year} {2018})},\ \bibinfo {note}
  {arXiv:1812.08223}\BibitemShut {NoStop}%
\end{thebibliography}%
\end{document}